\documentclass{article}

\PassOptionsToPackage{numbers,compress}{natbib}

\usepackage[preprint]{neurips_2026}

\usepackage[utf8]{inputenc}
\usepackage[T1]{fontenc}
\usepackage{microtype}
\usepackage{graphicx}
\usepackage{subcaption}
\usepackage{booktabs} 
\usepackage{wrapfig} 

\usepackage{hyperref}
\usepackage{url}
\usepackage{nicefrac}
\usepackage[dvipsnames]{xcolor}

\definecolor{gcyellow}{HTML}{fbc3aa}
\definecolor{gcred}{HTML}{ff6f79}

\newcommand{\ULTRAG}{\textsc{UltRAG }}
\newcommand{\ULTRAGnospace}{\textsc{UltRAG}}
\newcommand{\ULTRA}{\textsc{Ultra }}
\newcommand{\ULTRAnospace}{\textsc{Ultra}}
\newcommand{\ULTRAQUERY}{\textsc{UltraQuery }}
\newcommand{\ULTRAQUERYnospace}{\textsc{UltraQuery}}

\newcommand{\Kheeran}[1]{}

\usepackage{amsmath}
\usepackage{amssymb}
\usepackage{mathtools}
\usepackage{amsthm}
\usepackage{tikz}
\usepackage{multirow, makecell}
\usepackage{algorithm}
\usepackage{algorithmic}
\usetikzlibrary{calc, positioning, arrows.meta, fit, backgrounds}
\usepackage{tcolorbox}

\newtcolorbox{keyinsight}{
  colback=gray!10,
  colframe=gray!50,
  arc=3mm,
  boxrule=0.5pt,
  left=4pt,
  right=4pt,
  top=4pt,
  bottom=4pt
}

\usepackage[capitalize,noabbrev, nameinlink]{cleveref}
\hypersetup{
  colorlinks=true,
  citecolor=black,
  linkcolor=black,
  urlcolor=blue
}

\theoremstyle{plain}

\theoremstyle{definition}

\theoremstyle{remark}

\usepackage[textsize=tiny]{todonotes}

\title{UltRAG: a Universal Simple Scalable Recipe for Knowledge Graph RAG}

\author{%
  Dobrik Georgiev \\
  Graphcore Research \\
  \And
  Kheeran K. Naidu \\
  Graphcore Research \\
  \And
  Alberto Cattaneo \\
  Graphcore Research \\
  \AND
  Federico Monti \\
  Graphcore Research \\
  \And
  Carlo Luschi \\
  Graphcore Research \\
  \And
  Daniel Justus \\
  Graphcore Research \\
}

\usepackage{amsmath,amsfonts,bm}

\def\eqref#1{equation~\ref{#1}}

\def\1{\bm{1}}

\def\vzero{{\bm{0}}}

\def\vh{{\bm{h}}}

\def\vx{{\bm{x}}}

\DeclareMathAlphabet{\mathsfit}{\encodingdefault}{\sfdefault}{m}{sl}
\SetMathAlphabet{\mathsfit}{bold}{\encodingdefault}{\sfdefault}{bx}{n}

\def\gA{{\mathcal{A}}}

\def\gD{{\mathcal{D}}}
\def\gE{{\mathcal{E}}}
\def\gF{{\mathcal{F}}}
\def\gG{{\mathcal{G}}}

\def\gI{{\mathcal{I}}}

\def\gL{{\mathcal{L}}}

\def\gP{{\mathcal{P}}}

\def\gR{{\mathcal{R}}}
\def\gS{{\mathcal{S}}}
\def\gT{{\mathcal{T}}}

\def\gV{{\mathcal{V}}}

\def\gX{{\mathcal{X}}}

\begin{document}

\maketitle

\begin{abstract}

Large language models (LLMs) frequently generate confident yet factually incorrect content when used for language generation (a phenomenon often known as \emph{hallucination}). Retrieval augmented generation (RAG) tries to reduce factual errors by identifying information in a knowledge corpus and putting it in the context window of the model. While this approach is well-established for document-structured data, it is non-trivial to adapt it for Knowledge Graphs (KGs), especially for queries that require multi-node/multi-hop reasoning on graphs. We introduce \ULTRAGnospace, a training-free KG-RAG recipe that combines LLM query generation, a fully inductive neural query executor, and LLM arbitration. This off-the-shelf composition achieves state-of-the-art results on Knowledge Graph Question Answering (KGQA) tasks without retraining the LLM or executor, while enabling language models to interface with Wikidata-scale graphs (116M entities, 1.6B relations) at comparable or lower costs. Our ablation studies indicate that these gains come from the full system design rather than from any single component. 

\end{abstract}




\section{Introduction}


Large language models (LLMs) have rapidly become a key source of
information 
and a valuable tool for natural language tasks. However, despite
impressive linguistic capabilities from scaling model size and training data,
LLMs remain notoriously unreliable in their factual correctness. In particular,
a frequent failure mode is overconfidently generating plausible, yet fabricated
content that is unsupported or contradicted by facts \citep{huang2025survey}.

Mitigating these so-called \emph{hallucinations} is a central challenge for LLM
research, potentially unlocking a wide range of applications that strongly rely
on trustworthy language models. Retrieval-augmented generation (RAG) has
emerged as a central strategy for grounding LLMs in external knowledge
sources 
by identifying and retrieving information relevant to a query from an unstructured corpus of documents \citep{lewis2020RAG,borgeaud2022improving,izacard2023atlas}. However,
not all real-world information is organized as documents. Knowledge Graphs
(KGs) such as Wikidata \citep{vrandevcic2014wikidata} or domain-specific KGs, often containing proprietary
enterprise data, represent a huge source of reliable, interpretable, and easily
updatable information, stored in a highly efficient way as (subject,
predicate, object) triples. 
%
%
Unfortunately, translating the key concepts of RAG to graph-structure data is not an obvious task, as relevant facts are often distributed across multiple
entities and relations within the graph, which need to be considered together to provide a complete picture of the available knowledge. While some approaches for using KGs to augment LLMs have been developed \cite{mavromatis2025gnn, li2025simple}, they often fail for more complex queries that require deeper understanding of logic \citep{cattaneo2025ground}, and at the same time they do not scale well to large KGs with hundreds
of millions or even billions of entities and relations. 

\paragraph{Main Contributions} We present \ULTRAGnospace, a general framework for information retrieval from KGs
that can be applied efficiently and effectively to arbitrary {\em web-scale
graphs} and can be implemented with off-the-shelf components {\em without
retraining} the modules involved. To the best of our knowledge, it is the first
successful off-the-shelf LLM plus neural-query-executor KGQA framework evaluated
both on controlled question-specific subgraphs and in realistic retrieval from
Wikidata-scale KGs. 
The key idea behind \ULTRAG is that a successful LLM-based Knowledge Graph Question Answering (KGQA) system needs to be robust to {\em both} LLM and KG imperfections (\cref{sec:ultrag}). 
Motivated by this intuition, we argue for using {\em neural} query-execution modules for interfacing with real-world KGs. 
In our experiments, we show that such executors achieve $\sim$16\% improvement on average over symbolic ones when used as a tool by LLMs ceteris paribus. 
In addition, our version of \ULTRAG implemented with foundational query execution modules (\ULTRAGnospace-OTS) achieves zero-shot state-of-the art results on various inductive KGQA benchmarks (even when compared to transductive approaches finetuned on the specific datasets), and it is able to effectively scale to KGs the size of Wikidata (116 million entities and 1.6 billions triples). Further ablations show that all the key system components are necessary, that no component alone can bring the same improvements and that the overall method remains stable under its main hyperparameter choices (\autoref{app:wikidata_neural_symbolic}, \autoref{app:arbitrator_ablation}, \autoref{app:relation_id_ablation}, and \autoref{app:hyperparameter_sensitivity}). These results highlight a research direction that appears underexplored in the literature so far.   

\begin{figure*}[t]
    \centering
    \resizebox{\textwidth}{!}{%
        \begin{tikzpicture}[
    >=Stealth,
    seed/.style={rectangle, rounded corners, draw, fill=SkyBlue, inner sep=3pt, font=\small},
    person/.style={rectangle, rounded corners, draw, fill=green!10, inner sep=3pt, font=\small},
    uni/.style={rectangle, rounded corners, draw, fill=Dandelion, inner sep=3pt, font=\small},
    other/.style={rectangle, rounded corners, draw, fill=gray!15, inner sep=3pt, font=\small},
    rel/.style={midway, fill=white, inner sep=0.5pt, font=\scriptsize},
    legend/.style={font=\scriptsize, inner sep=0pt},
    highlight/.style={very thick, draw=BrickRed} 
]

\begin{scope}[yshift=-2mm]

\node[seed] (ta) {Q189};
\node[seed, below of=ta] (dl) {Q192};
\node[other, below of=dl] (cv) {Q244};

\node[other, above right=-2mm and 8mm of ta] (ra) {Q119};
\node[other, below left=5mm and -6mm of ra] (rc) {Q998};
\node[other, below=1mm of rc] (rb) {Q854};

\node[uni, below right=1mm and 4mm of ra]  (ux) {Q1009};
\node[uni, below left=4mm and -8mm of ux]  (uy) {Q5446};
\node[other, below right=9mm and -1mm of rc] (uz) {Q7611};

\draw[<-] (ta) to[bend left=10] node[rel,above=1mm] {P1} (ra);
\draw[<-, highlight] (ta) to[bend left=15] node[rel,above=1mm] {P1} (rc);

\draw[<-, highlight] (dl) -- node[rel,above=1.5mm] {P2} (rc);
\draw[<-] (dl) to[bend right=15] node[rel,above=0.5mm] {P2} (rb);

\draw[->] (cv) to[bend right=10] node[rel, below=1mm] {P6} (rb);

\draw[->] (ra) to[bend left=20] node[rel,above=1mm] {P4} (ux);
\draw[->,highlight] (rc) to[bend left=10] node[rel,above=1.0mm] {P4} (ux);
\draw[->,highlight] (rc) to[bend left=15] node[rel,above] {P4} (uy);
\draw[->] (rb) to[bend right=45] node[rel,above] {P5} (uz);

\draw[<->] (ra) to[bend right=20] node[rel,right=1mm] {P3} (rc);

\node[legend, align=left, above right=-1mm and 3mm of ux] (legE)  {
    \textbf{Q189}: Turing Award\\
    \textbf{Q192}: Deep Learning\\
    $\dots$
};
\node[legend, align=left, below=8mm of legE] (legR) {
    \textbf{P1}: Win\\
    \textbf{P2}: Field of work\\
    \textbf{P3}: Collaborates\\
    \textbf{P4}: University\\
    $\dots$
};

\begin{scope}[on background layer]
\node[
    draw,
    rounded corners,
    dashed,
    inner sep=4mm,
    fit=(ta) (dl) (cv) (ra) (rc) (rb) (ux) (uy) (uz) (legE) (legR)
] (KGBox) {};
\end{scope}
\node[fill=white, inner sep=0.5pt]
    at ($(KGBox.north west)!0.6!(KGBox.north)$) {\textbf{\textcolor{BrickRed}{Neural execution} on KG}};

\begin{scope}[on background layer]
\node[
    draw,
    rounded corners,
    dashed,
    inner sep=2mm,
    fit=(legE)
] (Ebox) {};
\end{scope}
\node[font=\scriptsize, fill=white, inner sep=0.5pt]
    at ($(Ebox.north west)!0.5!(Ebox.north)$) {\textbf{Entities}};

\begin{scope}[on background layer]
\node[
    draw,
    rounded corners,
    dashed,
    inner sep=2mm,
    fit=(legR)
] (Rbox) {};
\end{scope}
\node[font=\scriptsize, fill=white, inner sep=0.5pt]
    at ($(Rbox.north west)!0.55!(Rbox.north)$) {\textbf{Relations}};

\end{scope}



\node[
    draw,
    rounded corners,
    align=left,
    font=\footnotesize,
    inner sep=3pt,
    fill=white,
    text width=60mm,
] (qtext) [above left=-20mm and 8mm of KGBox] {%
    \textbf{Context:} Rules, relation types(+seed entities)\\
    \textbf{User:} In which universities do the
    Turing \\ Award winners of deep learning work in?\\
    \textbf{LLM:\ }{\ttfamily\scriptsize
    AND(
    \textcolor{SkyBlue}{Turing Award} -> P1\_inv,\\
    \ \ \quad\qquad\qquad\textcolor{SkyBlue}{Deep Learning} -> P2\_inv
    ) -> P4
    }
};


\node[above=2.8mm of qtext.south] (e1start) {};
\node[above=-1.0mm of qtext.south] (e2start) {};
\node[below left=3mm and 9mm of qtext.east] (betae) {};
\draw[->, thick, dashed, color=SkyBlue] (e1start) to[bend right=9] (ta) ;
\draw[->, thick, dashed, color=SkyBlue] (e2start) to[bend right=5] (dl) ;
\draw[->, thick, dashed, color=BrickRed] (betae) to ($(KGBox.west)+(0,1.0)$) ;


\node[
    draw,
    rounded corners,
    font=\footnotesize,
    inner sep=3pt,
    fill=white,
    text width=60mm,
    align=left,
] (exectext) [below=2mm of qtext.south] {%
    \textbf{Executor:} Top scoring entities are {\scriptsize\textcolor{Dandelion}{University\\ of Montreal (P=0.99)} $\dots$ RI School of Design (P=0.01)}
};

\node[
    draw,
    rounded corners,
    font=\footnotesize,
    inner sep=3pt,
    fill=white,
    text width=60mm,
    align=left,
] (llmresponse) [below=2mm of exectext.south] {%
    \textbf{LLM:} The universities are {\scriptsize\{Univ.\ of Montreal, Univ.\ of Toronto, $\dots$\}}
};

\begin{scope}[on background layer]
\draw[->, thick, dashed, color=Dandelion] ($(ux.south)!0.3!(ux.east)$) to[bend left=45] (exectext.east);
\draw[->, thick, dashed, color=Dandelion] ($(uy.south)!0.3!(uy.east)$) to[bend left=45] (exectext.east);
\end{scope}

\begin{scope}[on background layer]
\node[
    draw,
    rounded corners,
    dashed,
    inner sep=3mm,
    fit=(qtext) (exectext) (llmresponse)
] (QBox) {};
\end{scope}
\node[font=\footnotesize, fill=white, inner sep=0.5pt]
    at ($(QBox.north west)!0.35!(QBox.north)$) {\textbf{Conversation}};

\node[font=\tiny, text=gray, rotate=90] at ($(exectext.north west)+(-0.18,0.8)$) {$\dots$repeat if necessary$\dots$};

\draw[->, thick, color=gray!60] (exectext.west) -- ++(-1mm,0) -- ++(0,25mm) -- ($(qtext.north)+(-9mm,1.2mm)$) to[out=0, in=160] ($(qtext.north)$);

\end{tikzpicture}%
    }
    \caption{
        \ULTRAG pipeline. The LLM is provided with
        the syntactic rules for queries and the relation types. Ground-truth
        \textcolor{SkyBlue}{seed entities} (Turing Award, Deep Learning, etc.) may be given, but, if not, an \textcolor{SkyBlue}{entity
        linking} step takes place. The generated query is then
        \textcolor{BrickRed}{neurally executed} against the knowledge graph
        (each node receives a probability to be an answer at this stage). The
        most likely \textcolor{Dandelion}{query answers} are fed back to the
        LLM, which weighs both the returned probabilities and the semantic
        meaning of entities, and produces a final answer set.
    }\label{fig:vis_abstract}
\end{figure*}

\paragraph{Related work} 
For graph-structured knowledge sources, prior RAG approaches can be generally grouped into four main categories: KG agent-based approaches,
path-based approaches, graph neural network (GNN)-based approaches, and query-based approaches.
KG agents \citep{sun2024thinkongraph, chen2024planongraph} are LLMs that are trained to reason over KGs, starting from a given seed entity and exploring the KG step by step until a likely answer is found.
Path-based approaches \citep{luo2024reasoning, luo2025graph} first retrieve a set of relevant paths (a chain of relations) between the \emph{seed entities} (entities mentioned in the question) and other entities, and then use an LLM to reason over these paths to select the most likely answer.
GNN-based approaches have been relatively sporadic in the literature so far, with \citet{mavromatis2025gnn, mavromatis2025byokg} being among the most prominent ones. These methods compute node/relation embeddings and use them to compute a score for each entity to be the answer.
Similarly, query-based approaches, which translate natural questions into structured executable queries, have received little attention in the literature to date, largely due to relatively poor performance \citep{das2021case, yu2022decaf0, mavromatis2025gnn}.
While not falling squarely in any of the previous categories, we should also highlight that there are approaches, like SubgraphRAG \citep{li2025simple}, that fuse together ideas from more than one of the above groups. 

Analogously to \citet{li2025simple}, our approach can be classified as a hybrid solution for KGQA systems, since (as we shall see) it combines the use of LLMs (for generating queries and reasoning over retrived information) and GNNs (for running queries over KGs) to produce the desired answers. 
Yet, it differs from prior art by explicitly requiring the GNN to be aligned with the behaviour of a symbolic query executor.
The idea of aligning (graph) neural networks with algorithms (a.k.a.\ neural algorithmic reasoning; NAR) has first been proposed by \citet{velickovic2019neural}, with further theoretical foundations developed by \citet{xu2021neural, xu2020what}.
While there has been success in applications of this class of models \citep{deac2021xlvin, yu2025primaldual, numeroso2024dual}, their use as a LLM tool (this work) has never been explored to the best of our knowledge. Notably, if viewed through the lens of NAR, this work corresponds with the largest-scale application of NAR to date, with the second being \citet{numeroso2024dual} (graphs up to 6M edges).



\section{Background}

\paragraph{Knowledge Graphs} A knowledge graph $\gG=(\gV,\gR,\gE)$ is a database 
of entities $\gV$, relation types $\gR$, and relations between entities
$\gE \subset \gV\times\gR\times\gV$. Relation triplets $(h, r, t)\in\gE$
consist of a head $h$, tail $t$ and relation type $r$. Usually, $|\gR| \lll |\gE|$.
Knowledge graphs are often \emph{incomplete}
\citep{paulheim2016knowledge} and \emph{contain redundant information}
\citep{akrami2020realistic}, mainly due to automated construction
\citep{zhong2023comprehensive}.


\paragraph{From Link Prediction to Relation Projection}

Due to the incompleteness of knowledge graphs (i.e.\ missing relations between entities), there has been growing interest in learnable solutions for solving link prediction problems \citep{bordes2013translating,
yang2014embedding, trouillon2016complex, dettmers2018convolutional}. Graph Neural Networks (GNNs) appeared in particular as a prominent class of approaches for \emph{inductive} KG-completion tasks, thanks to their ability to generalize over unseen entities and connectivity patterns
\citep{schlichtkrull2018modeling, zhang2020learning, zhu2021neural, zhu2023net}.
In the domain of graph-based machine learning, the standard recipe for link
prediction uses the embeddings of the edge (relation) $\vh_r$, and
its two endpoints $\vh_u$ and $\vh_v$ \citep{bronstein2021geometric}, to predict the existence of a link. However,
in the presence of symmetries in the connectivity of knowledge graphs, this
recipe can underperform without the use of labelling tricks. As shown by 
\citet{zhang2021labeling}, when predicting whether $(u,r,v)$ exists, labelling
$u$ and $v$ differently from the remaining part of $\gV$ is theoretically optimal. Unfortunately, this also comes at high computational cost, as node embeddings need to be recomputed for each possible link one might want to predict. To address this limitation, ID-GNN \citep{you2021identity} reformulates the task, conditioning node embeddings {\em only} on source node $u$ (rather than on candidate link $(u,r,v)$), and then predicting the probability for {\em all} $v\in\gV$ to be the tail of an $r$ relation. Thanks to its reduced complexity, this formulation gained popularity in recent years, and it has been adopted by models such as NBFNet \cite{zhu2021neural}, which often serves as a baseline for inductive link prediction tasks.

\paragraph{Foundation models for knowledge graphs} While the previous paragraph discussed methodologies capable of generalizing over unseen nodes and connectivity patterns, an ideal link predictor should also process KGs with previously unseen relation types. In this direction, \citet{galkin2023towards} recently introduced \ULTRAnospace, a foundation model for knowledge graph completion, which generalizes across KGs with different relation vocabularies by making relation type embeddings a function of their relative interactions. To this end, \ULTRA builds an auxiliary graph $\gG_\gR=(\gR,\gR_{fund}, \gE_\gR)$,
with relation types $\gR$ as nodes, relation interactions as edges
$\gE_\gR$ and four fundamental relation interaction types
$\gR_{fund}=\{h2h, t2t, h2t, t2h\}$. These describe the four possible ways two relation types can interact (sharing the same head or tail, or the head/tail of one being the tail/head of the other). Whenever a relation projection $(u, r, ?)$ (i.e. a link prediction task) is performed on $\gG$, relation type embeddings are computed by applying a GNN on $\gG_\gR$. Tail probabilities for projection $(u, r, ?)$ are then inferred by initializing relation representations (and the embedding of source node $u$) on $\gG$ with the output of said GNN, and applying an inductive link predictor such as those described above. For further details, we refer the reader to
\citet{galkin2023towards}.



\paragraph{Foundation models for logical query reasoning} Beyond knowledge graph completion, 
another relevant task on KGs is complex logical query answering (CLQA). In short, provided a compositional query built by several projections and logical operators (i.e. written in first-order logic), the goal is to retrieve the set of entities that verify the query based on the information contained in the knowledge base. 
In the absence of noise, a symbolic executor would always be able to infer the correct entities that verify the query simply by ``walking" over the edges of the provided graph. However, in real-world scenarios, things are not as easy. Due to the incompleteness of the knowledge base, some relevant connections might indeed be missing, 
and as a result one might need to resort to link prediction modules to infer the likelihood of a node being part of the solution set. 
Along these lines and building on top of \ULTRAnospace, \citeauthor{galkin2024foundation} introduced \ULTRAQUERY \citep{galkin2024foundation}, a foundational logical query answering system operating with fuzzy sets\footnote{A fuzzy set $\gS$ is a generalization of a set where each element $u$ has "membership" $\mu_\gS(u)\in\left[0; 1\right]$ to $\gS$.}. 
Starting
from the leaves (e.g. \texttt{Turing award} in \cref{fig:vis_abstract}),
\ULTRAQUERY builds the answer upwards: it either projects an intermediate
result with a relation (e.g. $(y, \texttt{University}, ?)$) using
\textsc{Ultra}'s link prediction capabilities, or uses fuzzy logic to combine
previous projection results (at each intermediate step, the likelihood of a node satisfying a portion of the query can thus be seen as the membership function of a fuzzy set). While the architecture used for \ULTRA remains the same,
\citeauthor{galkin2024foundation} retrain its weights to make the model able to deal with non-leaf relation projections, where the input is a generic membership function over the node set, rather than a Kronecker's delta centered over a source node.


\section{\ULTRAG}
\label{sec:ultrag}

Our approach builds on the following two key insights:
\begin{keyinsight}
    \textbf{Insight \#1:} A successful query executor should be neural to handle
    LLM+KG noise.
\end{keyinsight}

The first insight comes from our observations that LLMs cannot be easily\footnote{E.g.\ giving relation types to the LLM or prompt
engineering.} and reliably constrained to build queries with \emph{only} existing
triplets (LLM noise). A method like GCR \citep{luo2025graph} that restricts the usable LLM output tokens for query generation requires a modification to the LLM pipeline, often infeasible and even impossible for closed LLMs. Moreover, KGs are incomplete (KG
noise) and have missing relations, so such an approach may be robust to LLM
noise but susceptible to KG noise.

Prior research has made similar observations \citep{das2021case,
yu2022decaf0, mavromatis2025gnn} which is why the field has mostly moved away from LLM-written graph queries run with symbolic executors, and instead delegates the query execution/reasoning to the LLM. This
brings us to our second insight:
\begin{keyinsight}
    \textbf{Insight \#2:} LLMs are \textbf{not} good neural executors.
\end{keyinsight}

LLMs underperform \citep{markeeva2024clrs, taylor2024large} on graph algorithm
simulation, especially the Bellman-Ford algorithm \citep{clrs2022}, when
tested on larger instances. This is not an expressivity issue for the
transformer architecture \citep{de2024simulation}, but to our best knowledge,
no LLM has reported robust performance on graph algorithms. This observation
is important in the context of graph-RAG, because SOTA link prediction approaches such as \ULTRA \cite{galkin2023towards} build on NBFNet
\citep{zhu2021neural}, which is a variation of the Bellman-Ford
algorithm. Even if LLMs would improve on their graph-reasoning capabilities in the upcoming years, LLM inference would remain significantly less efficient: a
back-of-the-envelope calculation, assuming near-perfect LLM conditions
(linear attention, one output token, etc.) would still take $10^6$ times more
FLOPs than lightweight models, such as graph neural networks -- see
\autoref{app:flops}. 
In light of these observations, we hypothesise that
{\em efficient and effective} query execution on KGs is better achieved with specialized neural query executors than with pure LLM-based approaches.



\begin{algorithm}[t]
\caption{\ULTRAG Recipe}
\label{alg:recipe}
\begin{algorithmic}[1]
\REQUIRE Question $q\in\gT$, Knowledge Graph $\gG=(\gV,\gR,\gE)$, Neural Query
Executor $\gX: 2^{\gF} \times \Phi \times \gG \to \gF$, Entity Linker
$\mathcal{L}: \Phi \times 2^{\gF} \to 2^{\gF}$, Decider
$\gD:\gF\times \gT \to \{\mathrm{True}, \mathrm{False}\}$, Arbitrator $\mathcal{A} :
\gF \times \Phi \times \gT \to 2^{\gV}$, \textcolor{ForestGreen}{(optional)} seed entities
$\gS \subseteq \gV$
\ENSURE Answer set $\gA \subseteq \gV$
\STATE $\mathcal{P} \gets
\begin{cases}
\{\vx_s : s \in \gS\}  & \text{if } \gS \text{ provided} \\
\emptyset & \text{otherwise}
\end{cases}$
\STATE $\textit{sufficient} \gets \text{False}$
\WHILE{$\neg\textit{sufficient}$}
    \STATE $\varphi \gets \text{LLM}(q, \gR, \gP)$ 
    \STATE $\gI \gets \textcolor{SkyBlue}{\mathcal{L}(\varphi, \gP)}$
    \STATE $\vx \gets \textcolor{BrickRed}{\gX(\gI, \varphi, \gG)}$
    \STATE $\textit{sufficient} \gets \gD(\vx, q)$
    \STATE $\gP \gets \gP\cup \{\vx\}$
\ENDWHILE
\STATE \textbf{return} \textcolor{Dandelion}{$\mathcal{A}(\vx, \varphi, q)$} 
\end{algorithmic}
\end{algorithm}


\textbf{Our framework is visualised in \cref{fig:vis_abstract} and described in detail in \cref{alg:recipe}}, with corresponding
parts colour coded for clarity. Apart from the LLM and the neural query executor $\gX$, the other key
components in our recipe are the entity linker $\gL$, the sufficiency decider $\gD$, and the arbitrator $\gA$. The algorithm operates over fuzzy sets with membership functions in $\gF = {[0,1]^{|\gV|}}$. It proceeds iteratively constructing queries and refining a set of partial answers $\gP$, until the membership function yielded by the neural query executor is deemed sufficient to answer the question. 
The partial answer set $\gP$ consits of fuzzy sets, and it is instantiated at the beginning of the loop with seed information $\gP = \{\vx_s= \1_{v = s}\}$ if available; otherwise $\gP$ is set to be the empty set. At each iteration,
\emph{after the query $\varphi$ is constructed}, the entity linker populates the leaves of the query with a set of membership functions $\gI=\{\vx_{l_1},\dots,\vx_{l_m}\}$, which connect the mentions associated to the leaves with the entities of the KG ($m$ is the number of leaves in the query). 
The result $\vx \in \gF$ of the query execution is then checked by the decider $\gD$ and the loop is terminated if $\vx$ is enough to answer the query. The arbitrator $\gA$ finally converts $\vx$ into the desired answer.

We strongly emphasise that our recipe is universal: we do not fix $\gL$,
$\gD$, $\gA$ or $\gX$, nor the logic used by the LLM. In fact,
\cref{sec:OTS} and \cref{sec:exp} show that our framework can be instantiated with only off-the-shelf tools. However, experiments suggest that parametrising $\gX$ as a neural network capable of simulating symbolic
reasoning on knowledge graphs is key to good performance. To our
best knowledge, this is the first use of a \emph{neural} query
executor in a KG RAG system.

\subsection{\ULTRAGnospace-OTS: An off-the-shelf \ULTRAG instance}\label{sec:OTS}



\paragraph{Simpler query language} Our first attempt used the BetaE-style
dataset format \citep{ren2020beta} from the original \ULTRAQUERY
implementation. Sadly, this tuple-based Domain Specific Language (DSL)
often produced heavy bracketing, hard to handle even for flagship LLMs like GPT-5
\citep{openai_gpt5_system_card_2025}. In our experiments, depending on the
dataset, $15$-$30\%$ of queries were invalid tuples and thus unexecutable. We
did not investigate this phenomenon, but we conjecture that it relates to
oversquashing \citep{barbero2024transformers} and attention sinks
\citep{gu2025when, arroyo2025bridging} in LLMs. 
We therefore developed a custom DSL ,
where projection only appends `\texttt{->}' plus the relation
identifier, and logical operators can be $n$-ary. This seemingly minor
edit led to almost no invalid queries. \autoref{app:query_language_comparison} provides further insights.

\paragraph{Entity linking} A notable example of the generality of our framework, via the fuzzy set parametrization, is that it allows for the non-existence of
uniquely identified seed entities (i.e. single entities that can be used as starting point for the reasoning process). 
Provided the target question, in the absence of seed entity information (i.e. $\gP = \emptyset$), the LLM can generate the structured query one needs to execute, together with mentions $\{l_1, \dots, l_m\}$ that will be matched against the knowledge base. Depending on similarity, each entity $v_j\in\gV$ receives a probability of being the seed entity. Specifically, in our experiments, the probabilities $p(v_j \text{ is seed entity of } l_i)$ are computed as:
\begin{equation}
d_{ij} = \|\mathrm{enc}(l_i) - \mathrm{enc}(v_j)\|_2, \quad
p(v_j \text{ is seed entity of } l_i) = \frac{\exp\left(-\frac{(d_{ij})^2}{2\sigma^2}\right)}{\sum_k \exp\left(-\frac{(d_{ik})^2}{2\sigma^2}\right)} \label{eq:RBF_norm}
\end{equation}

where 
$\sigma = 0.1$. 
The encoding function enc can be implemented with any pre-trained text embedding model (in our experiments we used E5$_{\mathrm{large}}$ \citep{wang2022text}), and embeddings for $v_i\in\gV$ can be inferred once and cached for multiple re-use. Given that knowledge bases can involve hundreds of millions of entities, efficient similarity search frameworks can additionally be used to enable \ULTRAG to deal with data of this scale. In our experiments, we used FAISS \citep{johnson2019billion, douze2024faiss}, specifically its inverted file with product quantization (IVFPQ; \citealp{jegou2011product}) approximate nearest neighbor search, to efficiently implement \cref{eq:RBF_norm}, and retrieve the $k$ most similar entities to a given mention.  


\paragraph{Query execution}
For what concerns the choice of the neural query executor, we use \ULTRAQUERY
\citep{galkin2024foundation} in our off-the-shelf implementation for its good
zero-shot performance and robustness to projection-operator choices. Using
relative relation type embeddings, in particular, allows the LLM to swap out a
relation type with its semantic equivalent (e.g. \texttt{Child} for
\texttt{Parent\_inv}), while achieving the same result from the query
execution. We expect improvements to knowledge graph
foundational models (e.g. \citep{huang2025expressive})
to translate to \textsc{UltRAG}, and relation projection methods with similar
foundational properties \citep{arun2025semma} to perform similarly. However,
we will not embark on a quantitative evaluation on the best choice of the neural query executor here -- our
aim is to show an instantiation of \ULTRAG that can \emph{efficiently} couple
LLMs with query execution, while achieving {\em state-of-the-art performance}. 


\ULTRAQUERY scales linearly with
the graph size both in time and in memory and can easily process all of \texttt{ogbl-wikikg2} (2.5M entities, 17M relations; \citealp{hu2020ogb}) on a
single GPU (GH200 96GB). 
While multi-GPU setups can in
principle be used for even larger KGs such as Wikidata, this would introduce architectural complexity and additional costs (both in terms of hardware and power consumption). 
Based on the observation that if answers exist for a generated query, they are typically located in a small neighborhood of the seed entities, we introduced a graph sampling step in our pipeline to reduce the amount of information that is fed in input to the query executor. Inspired by previous works in the graph machine learning literature \citep{klicpera2019predict, gasteiger2019diffusion, frasca2020sign}, we resorted in particular to personalised page rank \citep{page1998PageRank} to extract a relevant subgraph localized around seed entities. Due to space limitations, we provide details of this algorithm in \autoref{app:seppr}.

We conclude that our approach adds privacy. \ULTRAG is efficient enough for
local deployment, even on large databases. Further, unlike prior methods
\citep{sun2024thinkongraph, chen2024planongraph, li2025simple}, the graph
connectivity itself is not directly exposed to the LLM, nor do we need access to
the LLM weights.

\paragraph{Sufficiency and Query Arbitration}
While multiple iterations can improve performance by breaking complex questions into iteratively executed sub-queries, a single query execution was generally sufficient with modern powerful LLMs (GPT-5). A multi-step variant improves F1 by about two points, especially on complex classes (\autoref{app:multistep_ablation}); we use the single-step version in the main experiments for simplicity and efficiency. Therefore, $\gD$ in \ULTRAGnospace-OTS always returns {\em True}, terminating the while loop of \cref{alg:recipe} after one iteration.

Finally, we implemented the arbitrator $\gA$ as an LLM (GPT-5) that takes as input the top ranked entities in $\vx$, with their probabilities, and is prompted to return the final answer set. This stage is not a substitute for retrieval: in at least $94\%$ of generated responses, the arbitrator selects one of the top-$50$ executor candidates, and deterministic probability thresholds are substantially weaker (\autoref{app:arbitrator_ablation}). Using an LLM at this stage can be particularly beneficial: it lets the overall approach handle questions that cannot be directly answered with first-order logic (e.g. counting or temporal queries), while also letting the LLM use its knowledge to correct mistakes made by the query executor.





\section{Experiments}\label{sec:exp}

We present experimental results addressing multiple \textbf{R}esearch \textbf{Q}uestions
(\cref{sec:results}), with dataset, baseline, and reproduction details
(\cref{sec:exp_setup}).

\subsection{Experimental Setup}
\label{sec:exp_setup}

\paragraph{KGQA Datasets and Knowledge Graphs} For evaluating \ULTRAG, we focus
on datasets where the ground-truth source of information is already in the form of a
knowledge graph. This allows to remove any evaluation bias originating from a
knowledge-graph construction step (e.g. from web or textual data).
In line with this, datasets such as SimpleQA \citep{wei2024measuring} or FACTS
\citep{cheng2025facts} were excluded in our experiments.
In a recent work, \citet{zhang2025diagnosing} additionally observed that the average
correctness rate in some prominent KGQA datasets is too low for reliable benchmarking.
Therefore, we decided to not focus 
on widely used but error-prone datasets such as WebQSP \citep[][$52\%$ correct; MIT license]{yih2016value} or CWQ
\citep[][$49\%$ correct; MIT license]{talmor2018web}, and to prioritise instead newer, programmatically verified, datasets:
KGQAGen-10K \citep[CC-BY-4.0]{zhang2025diagnosing} and GTSQA \citep[CC-BY-4.0]{cattaneo2025ground}. For the
knowledge graph we use the Wikidata graph from \texttt{ogbl-wikikg2} (CC0) for GTSQA, and the \texttt{20251202} dump of Wikidata (CC0), processed as in
\citet{chen2024planongraph}\footnote{\url{https://github.com/liyichen-cly/PoG}; Apache-2.0.}, for KGQAGen-10K. For completeness, we also report WebQSP and CWQ results in \autoref{app:webqsp_cwq_results}. 

\paragraph{Baselines} We exclude methods constructing a knowledge graph
from another data format or retrieving documents as context, based on some
knowledge graph connectivity \citep{luo2025gfmrag}. Here we
compare \ULTRAGnospace-OTS against \emph{KG Agents}
\citep[ToG, Apache-2.0][]{sun2024thinkongraph},
\emph{path-based} approaches \citep[RoG/GCR, MIT][]{
luo2024reasoning, luo2025graph}, \emph{GNN-based} approaches \citep[GNN-RAG][]{mavromatis2025gnn} and {\em hybrid} approaches \citep[SubgraphRAG, MIT][]{li2025simple}.
We follow established literature in our choice of metrics. For ranking of
entities, we  report ranking-based scores -- Mean Reciprocal Rank (MRR) and
Hit@1,3,10. For final, discrete set answers, we report exact match hits, recall
and F1 of precision and recall. Unlike recall, Hits measures the
percentage of questions with at least one correct answer getting retrieved.


\paragraph{Training and Hyperparameters} We do not perform any training
on $\gX, \gA, \gD$. When ground truth entities are available, no training of $\gL$ is required. When they are not, a training phase is required with FAISS for the construction of the IVFPQ index (which takes less than 1 hr on a 64-core ARM Neoverse-V2). This makes
\ULTRAGnospace-OTS an inductive approach, as opposed to baselines, which
require retraining for each dataset. For \ULTRAQUERY (i.e. $\gX$), we used the
official
checkpoint\footnote{\url{https://github.com/DeepGraphLearning/ULTRA/tree/ultraquery}; MIT.} 
that is trained on FB15k-237, WN18RR, CoDEx-Medium. 

With FAISS, we used
$16,384$ cluster centroids for IVF, and $128$ 8-bit subquantizers per vector
embedding for PQ. For \cref{alg:ppr}, we picked $k=30,000$ for
both GTSQA and KGQAGen-10K, and passed the top 50 candidates to $\gA$, as ranked
by $\gX$. These choices are not brittle: a sensitivity analysis over the number of PPR nodes and the number of executor candidates passed to the arbitrator changes F1 by less than two points, and the best baseline remains about $21$ F1 points behind even under the weakest setting in this grid (\cref{tab:main_comparison_wikidata} and \autoref{app:hyperparameter_sensitivity}).

We should point out that, in our experiments, all our queries are constructed using only AND operators. While we could have added other logical
operators (OR, NOT, etc.), or allow for cyclic queries
\citep{cucumides2024unravl}, we found it unnecessary for obtaining good
performance. 
We also allow the LLM to use inverse projections by
adding an \texttt{\_inv} suffix. This can be done regardless of whether inverses
exist in the knowledge graph or not (e.g.\ \texttt{Child\_inv} is semantically equivalent to \texttt{Parent}). 
Lastly, we provide relation labels/descriptions for relation selection, but require the generated query to refer to relations by their identifiers
(\texttt{P<digits>} in Wikidata / WikiKG2). These identifiers are arbitrary handles rather than semantic features: randomly reassigning relation IDs leaves performance unchanged (\autoref{app:relation_id_ablation}).

\begin{figure*}[t]
    \centering
    \includegraphics[width=\textwidth]{./figures/special_combined.png}
    \caption{
        Comparison of \ULTRAQUERY vs symbolic query execution on GTSQA. Both receive identical queries generated by
        the LLM. Number inside brackets denotes projections, concatenation of
        expression denotes intersection. Best viewed on screen.
    }
    \label{fig:neural_vs_symbolic}
\end{figure*}
\subsection{Results}

\label{sec:results}

\paragraph{RQ1: How much does neural query execution improve a LLM’s ability to interface with KGs?} We compare the performance of \ULTRAQUERY against a symbolic executor, both receiving structured queries generated by GPT-5. We run both methods on the WikiKG2 subgraphs provided in GTSQA and rank the entities by their probability to be in the answer set (for symbolic execution each entity receives score 1 or 0). In \cref{fig:neural_vs_symbolic} we compare MRR and hit@10 for WikiKG2. \ULTRAQUERY consistently outperforms the query executor across all considered answer classes, achieving a class average improvement of $18.58\%$ (MRR) and $24.09\%$ (hit@10).
Due to space constraints, additional metrics and Wikidata results on 30,000-node \cref{alg:ppr} subgraphs are presented in \autoref{app:wikidata_neural_symbolic}. 

Taken together, the results highlight the benefit of using a {\em neural} query executor to make our framework resilient to both LLM and KG noise. Additional experiments comparing the performance  obtained with \ULTRAQUERY using noiseless queries generated by an oracle, to the ones produced by the LLM can be found in the Appendix in \cref{tab:query_type_results}. These highlight that, while GPT-5
is generally able to produce queries that correctly retrieve the answer nodes, this is by no means perfect, and further improvements in the LLM's query generation capabilities can only be expected to improve the performance discussed in the next paragraphs.
\begin{table}[t]
\centering
\caption{Comparison on question-specific graphs, for GTSQA (WikiKG2) and KGQAGen-10k (Wikidata). 
} \label{tab:main_comparison}
\begin{tabular}{lcccccc}
\toprule
& \multicolumn{3}{c}{\textbf{GTSQA}} & \multicolumn{3}{c}{\textbf{KGQAGen-10k}} \\
\cmidrule(lr){2-4} \cmidrule(lr){5-7}
\textbf{Model} & \textbf{Hits} & \textbf{Recall} & \textbf{F1} & \textbf{Hits} & \textbf{Recall} & \textbf{F1} \\
\midrule
    GPT-5-mini & 33.42 & 31.92 & 31.82 & 62.18 & 59.17 & 59.58 \\
    GPT-4.1 & 33.72 & 32.63 & 32.36 & 56.21 & 53.38 & 53.93 \\
    GPT-5 & 46.36 & 44.40 & 44.16 & 67.96 & 65.08 & 65.27 \\
\midrule
    ToG & 64.73 & 61.99 & 62.06 & 80.97 & 75.03 & 76.10 \\
\midrule
    GCR & 60.83 & 58.41 & 58.15 & 86.11 & 79.89 & 80.22 \\
    RoG & 76.51 & 74.61 & 73.99 & 88.43 & 84.69 & 84.92 \\
\midrule
    GNN-RAG & 76.76 & 74.95 & 74.90 & 82.56 & 77.98 & 78.92 \\
\midrule
    SubgraphRAG (200) & 84.34 & 81.66 & 81.62 & 89.76 & 85.52 & 86.28 \\
\midrule
\textbf{\ULTRAGnospace-OTS} & \textbf{92.66} & \textbf{91.05} & \textbf{89.29} & \textbf{92.04} & \textbf{90.77} & \textbf{88.82}\\
\bottomrule
\end{tabular}
\end{table}
\paragraph{RQ2: How does \ULTRAGnospace-OTS compare in performance against other KG-RAG approaches?}
To answer RQ2, we break down evaluation in two different parts. First, we compare \ULTRAGnospace-OTS and our baselines on the WikiKG2 subgraphs provided in GTSQA \cite{cattaneo2025ground} and on 
subgraphs of similar size constructed from Wikidata via PPR for KGQAGen-10K \cite{zhang2025diagnosing} (\cref{tab:main_comparison}). For both datasets, proof edges and answer nodes are always contained in the considered subgraphs, adding them back if they were lost during subsampling. This setting represents a simplified controlled version of a real-world scenario, where all the information a model needs to answer a given question is guaranteed to be available in the considered subgraph.
%
On GTSQA, \ULTRAG achieves $92.66\%$ exact match hits, $91.05\%$ recall, and $89.29\%$ F1. This represents an improvement of $+8.32\%$ in hits, $+9.39\%$ in recall, and $+8.03\%$ in F1 over the second best performing baseline (SubgraphRAG with 200 retrieved triples \cite{li2025simple}). On KGQAGen-10k, \ULTRAG achieves $92.04\%$ hits, $90.77\%$ recall, and $88.82\%$ F1, outperforming  the best baseline (again SubgraphRAG (200)) by respectively $+2.28\%$, $+5.25\%$, and $+2.54\%$. In the second part of our evaluation, we compare performance of the full pipeline described for \ULTRAGnospace-OTS against similar end-to-end pipelines implemented for RoG, GNN-RAG and SubgraphRAG (200) (the three best performing baselines in \cref{tab:main_comparison}). For these baselines, we use a LLM (GPT-5) to retrieve mentions that the LLM believes to be relevant for answering the provided question. We then link said mentions with entities in the provided KGs and extract relevant subgraphs in the same way we do for \ULTRAGnospace-OTS. \cref{tab:main_comparison_wikidata} provides a comparison of this evaluation for both transductive and inductive reasoning settings, together with results for an intermediate step, where we assume seed entities to be given. The entity-linking columns are the fully end-to-end setting: only the question is given, retrieval starts from the full KG, and no proof edges or answers are guaranteed in the extracted subgraph. The absolute scores are lower than in the ground-truth seed setting because the executor receives fuzzy entity-linking outputs rather than the crisp seed entities on which \ULTRAQUERY was trained, making this a harder out-of-distribution use of the executor. \ULTRAGnospace-OTS still improves F1 over the strongest baseline by $12.07$, $9.75$, and $1.36$ points on GTSQA (WikiKG2), GTSQA (Wikidata), and KGQAGen-10k, respectively.  

\begin{table*}[t]
\centering
\caption{Comparison on PPR subgraphs extracted from 
the ground-truth seed entities and seed entities obtained via Entity Linking. PPR subgraphs contain up to 30,000 nodes. All models use GPT-5 as reasoning LLM. 
We color-code settings that require {\bf \color{Purple} transductive} and {\bf \color{MidnightBlue} inductive} reasoning for the considered baselines; for GTSQA (Wikidata), we used baseline models originally trained on GTSQA (WikiKG2). We leave a `-' where a baseline cannot be applied. 
} \label{tab:main_comparison_wikidata}
\resizebox{\textwidth}{!}{
\begin{tabular}{lcccccccccccccccccc}
\toprule
& \multicolumn{9}{c}{\textbf{Ground-truth seed nodes}} & \multicolumn{9}{c}{\textbf{Entity Linking}} \\
\cmidrule(lr){2-10} \cmidrule(lr){11-19}
&  \multicolumn{3}{c}{\color{Purple} \textbf{GTSQA (WikiKG2) }} &  \multicolumn{3}{c}{\color{MidnightBlue}\textbf{GTSQA (Wikidata)}} & \multicolumn{3}{c}{\color{Purple} \textbf{KGQAGen-10k}} &  \multicolumn{3}{c}{\color{Purple}\textbf{GTSQA (WikiKG2)}} & \multicolumn{3}{c}{\color{MidnightBlue}\textbf{GTSQA (Wikidata)}} & \multicolumn{3}{c}{\color{Purple}\textbf{KGQAGen-10k}} \\
\cmidrule(lr){2-4} \cmidrule(lr){5-7} \cmidrule(lr){8-10} \cmidrule(lr){11-13} \cmidrule(lr){14-16} \cmidrule(lr){17-19}
\textbf{Model} & \textbf{Hits} & \textbf{Recall} & \textbf{F1} & \textbf{Hits} & \textbf{Recall} & \textbf{F1} & \textbf{Hits} & \textbf{Recall} & \textbf{F1} & \textbf{Hits} & \textbf{Recall} & \textbf{F1} & \textbf{Hits} & \textbf{Recall} & \textbf{F1} & \textbf{Hits} & \textbf{Recall} & \textbf{F1} \\
\midrule

    RoG & 72.63 & 70.73 & 69.81 & 62.55 & 59.81 & 58.93 & 86.25 & 82.60 & 82.68 & 72.00 & 69.97 & 68.88 & 60.64 & 57.66 & 56.85 & 82.42 & 78.89 & 79.22 \\
    GNN-RAG & 64.98 & 62.89 & 62.70 & - & - & - & 81.04 & 76.66 & 77.56 & 51.91 & 49.36 & 49.59 & - & - & - & 72.89 & 69.04 & 69.53 \\
    SubgraphRAG (200) & 73.98 & 71.14 & 70.91 & 63.29 & 59.60 & 59.56 & 80.47 & 75.88 & 76.54 & 71.82 & 69.37 & 69.01 & 58.61 & 55.43 & 55.04 & 76.68 & 72.63 & 73.17 \\
    \textbf{\ULTRAGnospace-OTS} & \textbf{90.81} & \textbf{89.39} & \textbf{87.18} & \textbf{86.74} & \textbf{84.47} & \textbf{82.08} & \textbf{90.62} & \textbf{89.41} & \textbf{87.63} & \textbf{85.70} & \textbf{83.71} & \textbf{81.08} & \textbf{72.93} & \textbf{70.48} & \textbf{66.60} & \textbf{83.98} & \textbf{82.59} & \textbf{80.58} \\
\bottomrule
\end{tabular}
}
\end{table*}

\paragraph{RQ3: How does \ULTRAGnospace-OTS perform with different LLMs (both of different families and sizes)?} For this particular research question we report in \cref{tab:ultrag_llm_ablation} (in \autoref{app:ultrag_llm_ablation} due to space constraints) an ablation study showing the performance achieved by \ULTRAGnospace-OTS with LLMs of the GPT and DeepSeek \cite{liu2025deepseek} families. For these experiments, we used PPR subgraphs extracted from Wikidata starting from ground truth seed entities. While using GPT-5 for both query generation and arbitration yields best performance (likely due to its reasoning capabilities), we note that all the tested combinations achieve good results in our analysis. Interestingly, the combination of GPT-5, for query generation, and GPT-5-mini, for arbitration, achieves close performance to our best model, suffering a reduction of only 4\% in Hits and F1 ($\sim$7\% for recall). These findings suggest that lighter LLMs could possibly be used in \ULTRAG whenever cost considerations are of relevance (especially for the arbitration stage, which likely requires less reasoning), while maintaining good performance overall.



\begin{wraptable}{r}{0.65\columnwidth}
\centering
\caption{Average efficiency per query on PPR subgraphs extracted for GTSQA from WikiKG2 (top) and Wikidata (bottom) with ground-truth seed nodes. All models use GPT-5 as reasoning LLM. Non-API running times are measured on GH200 chips, excluding API calls and PPR computation.} \label{tab:efficiency_comparison_gtsqa}
\resizebox{\linewidth}{!}{
\begin{tabular}{llccccc}
\toprule
\multirowcell{2}{\textbf{KG}} & \multirowcell{2}{\textbf{Model}} & \multirowcell{2}{\textbf{API cost}} & \textbf{\multirowcell{2}{Non-API \\ time (s)}} & \textbf{\multirowcell{2}{Input \\ tokens}} & \textbf{\multirowcell{2}{Output \\ tokens}} & \textbf{\multirowcell{2}{Cache hit \\ \%}} \\
& & & & & \\
\midrule
\multirowcell{4}{WikiKG2}
                                  & RoG  & 0.011\$ & $1.9 \pm 0.5$ & 0.9K & 2.2K & 0\\
                                  & GNN-RAG  & 0.012\$ & $9.9 \pm 2.3$ & 1.5K & 2.1K & 0\\
                                  & SubgraphRAG (200)  & 0.012\$ & $16.7 \pm 3.8$ & 2.7K & 2.1K & 0 \\
                                  & \textbf{\ULTRAGnospace-OTS} & 0.014\$ & $0.10 \pm 0.0$ & 23K & 2.2K & 93.77\\
\midrule
\multirowcell{3}{Wikidata}
  & RoG  & 0.013\$ & $4.0 \pm 1.1$ & 1.1K & 2.5K & 0 \\
  & SubgraphRAG (200) & 0.014\$ & $20.9 \pm 3.5$ & 3.1K & 2.4K & 0 \\
  &  \textbf{\ULTRAGnospace-OTS} & 0.017\$ & $0.15 \pm 0.1$ & 69K & 2.3K & 95.99\\
\bottomrule
\end{tabular}
}
\end{wraptable}

\paragraph{RQ4: How does \ULTRAGnospace-OTS compare with previous KG-RAG approaches in terms of efficiency?} To assess the efficiency of \ULTRAGnospace-OTS, we measured the average API cost per query with our LLM of choice (GPT-5), the average (non-API) runtime, and the average number of input/output tokens processed/generated by the LLM for the top-3 baselines of \cref{tab:main_comparison} and \ULTRAGnospace-OTS. All experiments start from the PPR subgraphs extracted from either WikiKG2 or Wikidata. 
We note that this comparison excludes entity linking and the subgraph extraction phases, as these are not defined for any of the considered baselines. A separate audit that adds PPR and FAISS retrieval time still leaves \ULTRAGnospace-OTS fastest, with $14.1\times$ and $4.3\times$ speed-ups on WikiKG2 and Wikidata, respectively (\autoref{app:retrieval_runtime}). 
Results for GTSQA are reported in \cref{tab:efficiency_comparison_gtsqa} and commented here, while results and analysis for KGQAGen are available in \autoref{app:runtime_costs}. On WikiKG2 subgraphs, 
\ULTRAGnospace-OTS is 19x/99x/167x faster than RoG/GNN-RAG/SubgraphRAG 
in terms of non-API time per query (similar results also hold for Wikidata subgraphs). Due to the presence of relation types in the prompt and two API calls, \ULTRAGnospace-OTS processes between 25x and 62x more input tokens than the baselines. However, $94-96\%$ of these tokens are cached by GPT-5 (due to our prompts being structurally identical), limiting both cost and computational complexity. The average number of output tokens is comparable across all methods. In terms of API cost, \ULTRAGnospace-OTS is 23\% to 27\% more expensive than the baselines (due to the larger size of the input), however, we emphasize that no additional cost would be required for extracting relevant mentions from input queries with \ULTRAGnospace-OTS, while all baselines would incur extra costs for this step in an end-to-end pipeline\footnote{With GPT-5 for mention extraction, baselines would incur an extra 0.003\$ (+490 outp.\ tok.) per query, making costs comparable.}.
Moreover, as shown in Table \ref{tab:ultrag_llm_ablation}, \ULTRAGnospace-OTS achieves a better performance than the baselines even when using smaller and cheaper LLMs, in which case \ULTRAGnospace-OTS compares favourably both in terms of results and efficiency.

\section{Conclusions}
In this work we introduced \ULTRAGnospace, a universal modular framework for Knowledge Graph Question Answering systems. Our off-the-shelf implementation, \ULTRAGnospace-OTS, achieves state-of-the-art results across a variety of settings and is able to effectively handle Wikidata-scale KGs in an end-to-end QA pipeline. Despite showing strong performance in standard KGQA settings, \ULTRAGnospace-OTS has limitations that may constrain its applicability for some real-world use cases. In particular, \ULTRAGnospace-OTS does not natively support temporal queries, which would require extending our methodology to Temporal Knowledge Graphs \cite{lin2023tflex}. 
Similarly, \ULTRAGnospace-OTS might underperform on Knowledge HyperGraphs due to its reliance on \ULTRAnospace, which has been shown to be suboptimal in this setting \cite{huang2025hyper}. Addressing these limitations constitutes an interesting research direction that we plan to investigate further in future work.

\paragraph*{Broader impact}
This paper presents work whose goal is to advance machine learning. There are many potential societal consequences of our work, none of which we feel must be highlighted.

\bibliography{example_paper}

@article{paulheim2016knowledge,
  title={Knowledge graph refinement: A survey of approaches and evaluation methods},
  author={Paulheim, Heiko},
  journal={Semantic web},
  volume={8},
  number={3},
  pages={489--508},
  year={2016},
  publisher={SAGE Publications Sage UK: London, England}
}

@inproceedings{akrami2020realistic,
  title={Realistic re-evaluation of knowledge graph completion methods: An experimental study},
  author={Akrami, Farahnaz and Saeef, Mohammed Samiul and Zhang, Qingheng and Hu, Wei and Li, Chengkai},
  booktitle={Proceedings of the 2020 ACM SIGMOD International Conference on Management of Data},
  pages={1995--2010},
  year={2020}
}

@article{zhong2023comprehensive,
  title={A comprehensive survey on automatic knowledge graph construction},
  author={Zhong, Lingfeng and Wu, Jia and Li, Qian and Peng, Hao and Wu, Xindong},
  journal={ACM Computing Surveys},
  volume={56},
  number={4},
  pages={1--62},
  year={2023},
  publisher={ACM New York, NY}
}

@article{zhu2021neural,
  title={Neural bellman-ford networks: A general graph neural network framework for link prediction},
  author={Zhu, Zhaocheng and Zhang, Zuobai and Xhonneux, Louis-Pascal and Tang, Jian},
  journal={Advances in neural information processing systems},
  volume={34},
  pages={29476--29490},
  year={2021}
}

@article{zhang2021labeling,
  title={Labeling trick: A theory of using graph neural networks for multi-node representation learning},
  author={Zhang, Muhan and Li, Pan and Xia, Yinglong and Wang, Kai and Jin, Long},
  journal={Advances in Neural Information Processing Systems},
  volume={34},
  pages={9061--9073},
  year={2021}
}

@inproceedings{schlichtkrull2018modeling,
  title={Modeling relational data with graph convolutional networks},
  author={Schlichtkrull, Michael and Kipf, Thomas N and Bloem, Peter and Van Den Berg, Rianne and Titov, Ivan and Welling, Max},
  booktitle={European semantic web conference},
  pages={593--607},
  year={2018},
  organization={Springer}
}

@article{zhu2023net,
  title={A* net: A scalable path-based reasoning approach for knowledge graphs},
  author={Zhu, Zhaocheng and Yuan, Xinyu and Galkin, Michael and Xhonneux, Louis-Pascal and Zhang, Ming and Gazeau, Maxime and Tang, Jian},
  journal={Advances in neural information processing systems},
  volume={36},
  pages={59323--59336},
  year={2023}
}

@article{bordes2013translating,
  title={Translating embeddings for modeling multi-relational data},
  author={Bordes, Antoine and Usunier, Nicolas and Garcia-Duran, Alberto and Weston, Jason and Yakhnenko, Oksana},
  journal={Advances in neural information processing systems},
  volume={26},
  year={2013}
}

@inproceedings{yang2014embedding,
  author       = {Bishan Yang and
                  Wen{-}tau Yih and
                  Xiaodong He and
                  Jianfeng Gao and
                  Li Deng},
  editor       = {Yoshua Bengio and
                  Yann LeCun},
  title        = {Embedding Entities and Relations for Learning and Inference in Knowledge Bases},
  booktitle    = {3rd International Conference on Learning Representations, {ICLR} 2015,
                  San Diego, CA, USA, May 7--9, 2015, Conference Track Proceedings},
  year         = {2015},
  url          = {http://arxiv.org/abs/1412.6575}
}

@inproceedings{trouillon2016complex,
  title={Complex embeddings for simple link prediction},
  author={Trouillon, Th{\'e}o and Welbl, Johannes and Riedel, Sebastian and Gaussier, {\'E}ric and Bouchard, Guillaume},
  booktitle={International conference on machine learning},
  pages={2071--2080},
  year={2016},
  organization={PMLR}
}

@inproceedings{dettmers2018convolutional,
  title={Convolutional 2d knowledge graph embeddings},
  author={Dettmers, Tim and Minervini, Pasquale and Stenetorp, Pontus and Riedel, Sebastian},
  booktitle={Proceedings of the AAAI conference on artificial intelligence},
  volume={32},
  number={1},
  year={2018}
}

@inproceedings{zhang2020learning,
  title={Learning hierarchy-aware knowledge graph embeddings for link prediction},
  author={Zhang, Zhanqiu and Cai, Jianyu and Zhang, Yongdong and Wang, Jie},
  booktitle={Proceedings of the AAAI conference on artificial intelligence},
  volume={34},
  number={03},
  pages={3065--3072},
  year={2020}
}

@article{bronstein2021geometric,
      title="{Geometric Deep Learning: Grids, Groups, Graphs, Geodesics, and Gauges}", 
      author={Michael M. Bronstein and Joan Bruna and Taco Cohen and Petar Veličković},
      year={2021},
      journal={arXiv preprint arXiv:2104.13478},
}

@inproceedings{you2021identity,
  title={Identity-aware graph neural networks},
  author={You, Jiaxuan and Gomes-Selman, Jonathan M and Ying, Rex and Leskovec, Jure},
  booktitle={Proceedings of the AAAI conference on artificial intelligence},
  volume={35},
  number={12},
  pages={10737--10745},
  year={2021}
}

@inproceedings{galkin2023towards,
  author       = {Mikhail Galkin and
                  Xinyu Yuan and
                  Hesham Mostafa and
                  Jian Tang and
                  Zhaocheng Zhu},
  title        = {Towards Foundation Models for Knowledge Graph Reasoning},
  booktitle    = {The Twelfth International Conference on Learning Representations,
                  {ICLR} 2024, Vienna, Austria, May 7--11, 2024},
  publisher    = {OpenReview.net},
  year         = {2024},
  url          = {https://openreview.net/forum?id=jVEoydFOl9}
}

@article{galkin2024foundation,
  title={A foundation model for zero-shot logical query reasoning},
  author={Galkin, Michael and Zhou, Jincheng and Ribeiro, Bruno and Tang, Jian and Zhu, Zhaocheng},
  journal={Advances in Neural Information Processing Systems},
  volume={37},
  pages={54137--54160},
  year={2024}
}

@inproceedings{huang2025expressive,
  author       = {Xingyue Huang and
                  Pablo Barcel{\'{o}} and
                  Michael M. Bronstein and
                  {\.I}smail {\.I}lkan Ceylan and
                  Mikhail Galkin and
                  Juan L. Reutter and
                  Miguel A. Romero Orth},
  title        = {How Expressive are Knowledge Graph Foundation Models?},
  booktitle    = {Forty-second International Conference on Machine Learning, {ICML}
                  2025, Vancouver, BC, Canada, July 13--19, 2025},
  publisher    = {OpenReview.net},
  year         = {2025},
  url          = {https://openreview.net/forum?id=mXEdUcLtaK}
}

@inproceedings{markeeva2024clrs,
  title={The clrs-text algorithmic reasoning language benchmark},
  author={Markeeva, Larisa and McLeish, Sean and Ibarz, Borja and Bounsi, Wilfried and Kozlova, Olga and Vitvitskyi, Alex and Blundell, Charles and Goldstein, Tom and Schwarzschild, Avi and Veli{\v{c}}kovi{\'c}, Petar},
  booktitle={ICML 2024 Workshop on Data-centric Machine Learning Research (DMLR): Datasets for Foundation Models},
  year={2024},
  note={Workshop poster (listed on ICML 2024 virtual site)},
  url={https://openreview.net/forum?id=cG0UutsUYh}
}

@article{taylor2024large,
  title={Are Large-Language Models Graph Algorithmic Reasoners?},
  author={Taylor, Alexander K and Cuturrufo, Anthony and Yathish, Vishal and Ma, Mingyu Derek and Wang, Wei},
  journal={arXiv preprint arXiv:2410.22597},
  year={2024}
}

@book{clrs2022,
  author    = {Thomas H. Cormen and Charles E. Leiserson and Ronald L. Rivest and Clifford Stein},
  title     = {Introduction to Algorithms},
  edition   = {4},
  publisher = {MIT Press},
  year      = {2022}
}

@misc{openai2025gptoss,
  title         = {gpt-oss-120b \& gpt-oss-20b Model Card},
  author        = {{OpenAI}},
  year          = {2025},
  month         = aug,
  eprint        = {2508.10925},
  archivePrefix = {arXiv},
  primaryClass  = {cs.CL},
  doi           = {10.48550/arXiv.2508.10925},
  url           = {https://arxiv.org/abs/2508.10925}
}

@inproceedings{
cucumides2024unravl,
title={UnRavL: A Neuro-Symbolic Framework for Answering Graph Pattern Queries in Knowledge Graphs},
author={Tamara Cucumides and Daniel Daza and Pablo Barcelo and Michael Cochez and Floris Geerts and Juan L Reutter and Miguel Romero Orth},
booktitle={The Third Learning on Graphs Conference},
year={2024},
url={https://openreview.net/forum?id=183XrFqaHN}
}

@article{ren2020beta,
  title={Beta embeddings for multi-hop logical reasoning in knowledge graphs},
  author={Ren, Hongyu and Leskovec, Jure},
  journal={Advances in Neural Information Processing Systems},
  volume={33},
  pages={19716--19726},
  year={2020}
}

@article{barbero2024transformers,
  title={Transformers need glasses! information over-squashing in language tasks},
  author={Barbero, Federico and Banino, Andrea and Kapturowski, Steven and Kumaran, Dharshan and Madeira Ara{\'u}jo, Jo{\~a}o and Vitvitskyi, Oleksandr and Pascanu, Razvan and Veli{\v{c}}kovi{\'c}, Petar},
  journal={Advances in Neural Information Processing Systems},
  volume={37},
  pages={98111--98142},
  year={2024}
}

@article{arroyo2025bridging,
  title={Bridging Graph Neural Networks and Large Language Models: A Survey and Unified Perspective},
  author={Arroyo, {\'A}lvaro and Barbero, Federico and Blayney, Hugh and Bronstein, Michael and Dong, Xiaowen and Li{\`o}, Pietro and Pascanu, Razvan and Vandergheynst, Pierre},
  year={2025},
  publisher={OpenReview}
}

@inproceedings{
gu2025when,
title={When Attention Sink Emerges in Language Models: An Empirical View},
author={Xiangming Gu and Tianyu Pang and Chao Du and Qian Liu and Fengzhuo Zhang and Cunxiao Du and Ye Wang and Min Lin},
booktitle={The Thirteenth International Conference on Learning Representations},
year={2025},
url={https://openreview.net/forum?id=78Nn4QJTEN}
}

@misc{openai_gpt5_system_card_2025,
  author       = {{OpenAI}},
  title        = {{GPT-5 System Card}},
  year         = {2025},
  month        = {August},
  howpublished = {\url{https://cdn.openai.com/gpt-5-system-card.pdf}},
  note         = {System card detailing GPT-5’s architecture, safety evaluations, and mitigations. Accessed December 30, 2025.}
}

@article{johnson2019billion,
  title={Billion-scale similarity search with {GPUs}},
  author={Johnson, Jeff and Douze, Matthijs and J{\'e}gou, Herv{\'e}},
  journal={IEEE Transactions on Big Data},
  volume={7},
  number={3},
  pages={535--547},
  year={2019},
  publisher={IEEE}
}

@article{douze2024faiss,
      title={The Faiss library},
      author={Matthijs Douze and Alexandr Guzhva and Chengqi Deng and Jeff Johnson and Gergely Szilvasy and Pierre-Emmanuel Mazaré and Maria Lomeli and Lucas Hosseini and Hervé Jégou},
      year={2024},
      eprint={2401.08281},
      archivePrefix={arXiv},
      primaryClass={cs.LG}
}

@article{wang2022text,
  title={Text embeddings by weakly-supervised contrastive pre-training},
  author={Wang, Liang and Yang, Nan and Huang, Xiaolong and Jiao, Binxing and Yang, Linjun and Jiang, Daxin and Majumder, Rangan and Wei, Furu},
  journal={arXiv preprint arXiv:2212.03533},
  year={2022}
}

@article{jegou2011product,
  title={Product Quantization for Nearest Neighbor Search},
  author={J{\'e}gou, Herv{\'e} and Douze, Matthijs and Schmid, Cordelia},
  journal={IEEE Transactions on Pattern Analysis and Machine Intelligence},
  volume={33},
  number={1},
  pages={117--128},
  year={2011}
}

@article{arun2025semma,
  title={SEMMA: A Semantic Aware Knowledge Graph Foundation Model},
  author={Arun, Arvindh and Kumar, Sumit and Nayyeri, Mojtaba and Xiong, Bo and Kumaraguru, Ponnurangam and Vergari, Antonio and Staab, Steffen},
  journal={arXiv preprint arXiv:2505.20422},
  year={2025}
}

@techreport{page1998PageRank,
  author = {Page, Larry and Brin, Sergey and Motwani, Rajeev and Winograd, Terry},
  title = {{The PageRank Citation Ranking: Bringing Order to the Web}},
  institution = {Stanford University},
  year = {1998},
  number = {SIDL-WP-1998-0120},
  url = {ilpubs.stanford.edu},
  note = {Often cited as Page et al. (1999) in academic literature}
}

@inproceedings{klicpera2019predict,
  title={Predict then Propagate: Graph Neural Networks meet Personalized PageRank},
  author={Klicpera, Johannes and Bojchevski, Aleksandar and G{\"u}nnemann, Stephan},
  booktitle={International Conference on Learning Representations (ICLR)},
  year={2019}
}

@inproceedings{
sun2024thinkongraph,
title={Think-on-Graph: Deep and Responsible Reasoning of Large Language Model on Knowledge Graph},
author={Jiashuo Sun and Chengjin Xu and Lumingyuan Tang and Saizhuo Wang and Chen Lin and Yeyun Gong and Lionel Ni and Heung-Yeung Shum and Jian Guo},
booktitle={The Twelfth International Conference on Learning Representations},
year={2024},
url={https://openreview.net/forum?id=nnVO1PvbTv}
}

@inproceedings{
chen2024planongraph,
title={Plan-on-Graph: Self-Correcting Adaptive Planning of Large Language Model on Knowledge Graphs},
author={Liyi Chen and Panrong Tong and Zhongming Jin and Ying Sun and Jieping Ye and Hui Xiong},
booktitle={The Thirty-eighth Annual Conference on Neural Information Processing Systems},
year={2024},
url={https://openreview.net/forum?id=CwCUEr6wO5}
}

@inproceedings{li2025simple,
  author    = {Mufei Li and Siqi Miao and Pan Li},
  title     = {Simple is Effective: The Roles of Graphs and Large Language Models in Knowledge-Graph-Based Retrieval-Augmented Generation},
  booktitle = {The Thirteenth International Conference on Learning Representations, {ICLR} 2025, Singapore, April 24-28, 2025},
  publisher = {OpenReview.net},
  year      = {2025},
  url       = {https://openreview.net/forum?id=JvkuZZ04O7},
  bibsource = {dblp computer science bibliography, https://dblp.org}
}

@article{wei2024measuring,
  title        = {Measuring Short-Form Factuality in Large Language Models},
  author       = {Wei, Jason and Nguyen, Karina and Chung, Hyung Won and Jiao, Yunxin Joy and Papay, Spencer and Glaese, Amelia and Schulman, John and Fedus, William},
  journal      = {arXiv preprint arXiv:2411.04368},
  year         = {2024},
  url          = {https://arxiv.org/abs/2411.04368}
}

@article{cheng2025facts,
  title        = {The {FACTS} Leaderboard: A Comprehensive Benchmark for Large Language Model Factuality},
  author       = {Cheng, Aileen and Jacovi, Alon and Globerson, Amir and Golan, Ben and Kwong, Charles and Alberti, Chris and Tao, Connie and Ben-David, Eyal and Singh Tomar, Gaurav and Haas, Lukas and Bitton, Yonatan and Bloniarz, Adam and Bai, Aijun and Wang, Andrew and Siddiqui, Anfal and Bajuelos Castillo, Arturo and Atias, Aviel and others},
  year         = {2025},
  journal      = {arXiv preprint arXiv:2512.10791},
  url          = {https://arxiv.org/abs/2512.10791}
}

@inproceedings{yih2016value,
  title        = {The Value of Semantic Parse Labeling for Knowledge Base Question Answering},
  author       = {Yih, Wen-tau and Richardson, Matthew and Meek, Christopher and Chang, Ming-Wei and Suh, Jina},
  booktitle    = {Proceedings of the 54th Annual Meeting of the Association for Computational Linguistics (Volume 2: Short Papers)},
  year         = {2016},
  pages        = {201--206},
  address      = {Berlin, Germany},
  publisher    = {Association for Computational Linguistics},
  doi          = {10.18653/v1/P16-2033}
}

@inproceedings{talmor2018web,
  title        = {The Web as a Knowledge-Base for Answering Complex Questions},
  author       = {Talmor, Alon and Berant, Jonathan},
  booktitle    = {Proceedings of the 2018 Conference of the North American Chapter of the Association for Computational Linguistics: Human Language Technologies (Long Papers)},
  year         = {2018},
  pages        = {641--651},
  address      = {New Orleans, Louisiana},
  publisher    = {Association for Computational Linguistics},
  doi          = {10.18653/v1/N18-1059}
}

@article{cattaneo2025ground,
  title        = {Ground-Truth Subgraphs for Better Training and Evaluation of Knowledge Graph Augmented LLMs},
  author       = {Cattaneo, Alberto and Luschi, Carlo and Justus, Daniel},
  journal      = {arXiv preprint arXiv:2511.04473},
  year         = {2025},
  url          = {https://arxiv.org/abs/2511.04473}
}

@inproceedings{
zhang2025diagnosing,
title={Diagnosing and Addressing Pitfalls in {KG}-{RAG} Datasets: Toward More Reliable Benchmarking},
author={Liangliang Zhang and Zhuorui Jiang and Hongliang Chi and Haoyang Chen and Mohammed ElKoumy and Fali Wang and Qiong Wu and Zhengyi Zhou and Shirui Pan and Suhang Wang and Yao Ma},
booktitle={The Thirty-ninth Annual Conference on Neural Information Processing Systems Datasets and Benchmarks Track},
year={2025},
url={https://openreview.net/forum?id=Vd5JXiX073}
}

@article{luo2025gfmrag,
  title={GFM-RAG: Graph Foundation Model for Retrieval Augmented Generation},
  author={Luo, Linhao and Zhao, Zicheng and Haffari, Gholamreza and Phung, Dinh and Gong, Chen and Pan, Shirui},
  journal={NeurIPS 2025},
  year={2025}
}

@inproceedings{luo2025graph,
  title={Graph-constrained Reasoning: Faithful Reasoning on Knowledge Graphs with Large Language Models},
  author={Luo, Linhao and Zhao, Zicheng and Haffari, Gholamreza and Li, Yuan-Fang and Gong, Chen and Pan, Shirui},
  booktitle={Forty-second International Conference on Machine Learning},
  year={2025}
}

@article{yu2022decaf0,
  title     = {DecAF: Joint Decoding of Answers and Logical Forms for Question Answering over Knowledge Bases},
  author    = {Donghan Yu and Shenmin Zhang and Patrick Ng and Henghui Zhu and A. Li and J. Wang and Yiqun Hu and William Wang and Zhiguo Wang and Bing Xiang},
  journal   = {International Conference on Learning Representations},
  year      = {2022},
  doi       = {10.48550/arXiv.2210.00063},
  bibSource = {Semantic Scholar https://www.semanticscholar.org/paper/0751b6de6c07c52a324450e32ae6581d403603da}
}

@inproceedings{das2021case,
  title={Case-based Reasoning for Natural Language Queries over Knowledge Bases},
  author={Das, Rajarshi and Zaheer, Manzil and Thai, Dung and Godbole, Ameya and Perez, Ethan and Lee, Jay-Yoon and Tan, Lizhen and Polymenakos, Lazaros and Mccallum, Andrew},
  booktitle={Proceedings of the 2021 Conference on Empirical Methods in Natural Language Processing},
  pages={9594--9611},
  year={2021}
}

@inproceedings{mavromatis2025gnn,
  title={Gnn-rag: Graph neural retrieval for efficient large language model reasoning on knowledge graphs},
  author={Mavromatis, Costas and Karypis, George},
  booktitle={Findings of the Association for Computational Linguistics: ACL 2025},
  pages={16682--16699},
  year={2025}
}

@inproceedings{de2024simulation,
  title={Simulation of graph algorithms with looped transformers},
  author={De Luca, Arturs Back and Fountoulakis, Kimon},
  booktitle={Proceedings of the 41st International Conference on Machine Learning},
  pages={2319--2363},
  year={2024}
}

@inproceedings{mavromatis2025byokg,
  title={Byokg-rag: Multi-strategy graph retrieval for knowledge graph question answering},
  author={Mavromatis, Costas and Adeshina, Soji and Ioannidis, Vassilis N and Han, Zhen and Zhu, Qi and Robinson, Ian and Thompson, Bryan and Rangwala, Huzefa and Karypis, George},
  booktitle={Proceedings of the 2025 Conference on Empirical Methods in Natural Language Processing},
  pages={27869--27886},
  year={2025}
}

@inproceedings{luo2024reasoning,
  author    = {Linhao Luo and Yuan{-}Fang Li and Gholamreza Haffari and Shirui Pan},
  title     = {Reasoning on Graphs: Faithful and Interpretable Large Language Model Reasoning},
  booktitle = {The Twelfth International Conference on Learning Representations, {ICLR} 2024, Vienna, Austria, May 7-11, 2024},
  publisher = {OpenReview.net},
  year      = {2024},
  url       = {https://openreview.net/forum?id=ZGNWW7xZ6Q},
}

@article{huang2025survey,
  title={A survey on hallucination in large language models: Principles, taxonomy, challenges, and open questions},
  author={Huang, Lei and Yu, Weijiang and Ma, Weitao and Zhong, Weihong and Feng, Zhangyin and Wang, Haotian and Chen, Qianglong and Peng, Weihua and Feng, Xiaocheng and Qin, Bing and Liu, Ting},
  journal={ACM Transactions on Information Systems},
  volume={43},
  number={2},
  pages={1--55},
  year={2025},
  publisher={ACM New York, NY}
}

@article{lewis2020RAG,
  title={Retrieval-augmented generation for knowledge-intensive nlp tasks},
  author={Lewis, Patrick and Perez, Ethan and Piktus, Aleksandra and Petroni, Fabio and Karpukhin, Vladimir and Goyal, Naman and K{\"u}ttler, Heinrich and Lewis, Mike and Yih, Wen-tau and Rockt{\"a}schel, Tim and others},
  journal={Advances in Neural Information Processing Systems},
  volume={33},
  pages={9459--9474},
  year={2020}
}

@article{izacard2023atlas,
  title={Atlas: Few-shot learning with retrieval augmented language models},
  author={Izacard, Gautier and Lewis, Patrick and Lomeli, Maria and Hosseini, Lucas and Petroni, Fabio and Schick, Timo and Dwivedi-Yu, Jane and Joulin, Armand and Riedel, Sebastian and Grave, Edouard},
  journal={Journal of Machine Learning Research},
  volume={24},
  number={251},
  pages={1--43},
  year={2023}
}

@inproceedings{borgeaud2022improving,
  title={Improving language models by retrieving from trillions of tokens},
  author={Borgeaud, Sebastian and Mensch, Arthur and Hoffmann, Jordan and Cai, Trevor and Rutherford, Eliza and Millican, Katie and Van Den Driessche, George Bm and Lespiau, Jean-Baptiste and Damoc, Bogdan and Clark, Aidan and others},
  booktitle={International conference on machine learning},
  pages={2206--2240},
  year={2022},
  organization={PMLR}
}

@article{velickovic2019neural,
  title={Neural Execution of Graph Algorithms},
  author={Veli{\v{c}}kovi{\'c}, Petar and Ying, Rex and Padovano, Matilde and Hadsell, Raia and Blundell, Charles},
  journal={International Conference on Learning Representations},
  year={2020},
  url={https://openreview.net/forum?id=SkgKO0EtvS}
}

@inproceedings{xu2021neural,
  title={How Neural Networks Extrapolate: From Feedforward to Graph Neural Networks},
  author={Xu, Keyulu and Zhang, Mozhi and Li, Jingling and Leskovec, Jure and Jegelka, Stefanie and others},
  booktitle={International Conference on Learning Representations},
  year={2021},
  url={https://openreview.net/forum?id=UH-cmocLJC}
}

@inproceedings{xu2020what,
  title={What Can Neural Networks Reason About?},
  author={Xu, Keyulu and Li, Jingling and Zhang, Muhan and Du, Simon S. and Kawarabayashi, Ken-ichi and Jegelka, Stefanie},
  booktitle={International Conference on Learning Representations},
  year={2020},
  url={https://openreview.net/forum?id=rJxbJeHFPS}
}

@inproceedings{deac2021xlvin,
  title={{XLVIN}: eXecuted Latent Value Iteration Nets},
  author={Deac, Andreea and Veli{\v{c}}kovi{\'c}, Petar and Milinkovi{\'c}, Ognjen and Bacon, Pierre-Luc and Tang, Jian and Li{\`o}, Pietro},
  booktitle={Third Workshop on Graphs and more Complex Structures for Learning and Reasoning at AAAI 2021},
  year={2021}
}

@inproceedings{
yu2025primaldual,
title={Primal-Dual Neural Algorithmic Reasoning},
author={Yu He and Ellen Vitercik},
booktitle={Forty-second International Conference on Machine Learning},
year={2025},
url={https://openreview.net/forum?id=iBpkzB5LEr}
}

@article{numeroso2024dual,
  title={Dual Algorithmic Reasoning},
  author={Numeroso, Danilo and Bacciu, Davide and Veli{\v{c}}kovi{\'c}, Petar},
  journal={The Twelfth International Conference on Learning Representations},
  year={2024},
  url={https://openreview.net/forum?id=bnEarBBf7q}
}

@article{vrandevcic2014wikidata,
  title={Wikidata: a free collaborative knowledgebase},
  author={Vrande{\v{c}}i{\'c}, Denny and Kr{\"o}tzsch, Markus},
  journal={Communications of the ACM},
  volume={57},
  number={10},
  pages={78--85},
  year={2014},
  publisher={ACM New York, NY, USA}
}

@article{frasca2020sign,
  title={Sign: Scalable inception graph neural networks},
  author={Frasca, Fabrizio and Rossi, Emanuele and Eynard, Davide and Chamberlain, Ben and Bronstein, Michael and Monti, Federico},
  journal={arXiv preprint arXiv:2004.11198},
  year={2020}
}

@article{gasteiger2019diffusion,
  title={Diffusion improves graph learning},
  author={Gasteiger, Johannes and Wei{\ss}enberger, Stefan and G{\"u}nnemann, Stephan},
  journal={Advances in neural information processing systems},
  volume={32},
  year={2019}
}

@article{huang2025hyper,
  title={HYPER: A Foundation Model for Inductive Link Prediction with Knowledge Hypergraphs},
  author={Huang, Xingyue and Galkin, Mikhail and Bronstein, Michael M and Ceylan, {\.I}smail {\.I}lkan},
  journal={arXiv preprint arXiv:2506.12362},
  year={2025}
}

@article{lin2023tflex,
  title={TFLEX: Temporal feature-logic embedding framework for complex reasoning over temporal knowledge graph},
  author={Lin, Xueyuan and Xu, Chengjin and Zhou, Gengxian and Luo, Haoran and Hu, Tianyi and Su, Fenglong and Li, Ningyuan and Sun, Mingzhi and others},
  journal={Advances in Neural Information Processing Systems},
  volume={36},
  pages={73039--73081},
  year={2023}
}

@article{hu2020ogb,
  title={Open Graph Benchmark: Datasets for Machine Learning on Graphs},
  author={Hu, Weihua and Fey, Matthias and Zitnik, Marinka and Dong, Yuxiao and Ren, Hongyu and Liu, Bowen and Catasta, Michele and Leskovec, Jure},
  journal={arXiv preprint arXiv:2005.00687},
  year={2020}
}

@article{liu2025deepseek,
  title={Deepseek-v3. 2: Pushing the frontier of open large language models},
  author={Liu, Aixin and Mei, Aoxue and Lin, Bangcai and Xue, Bing and Wang, Bingxuan and Xu, Bingzheng and Wu, Bochao and Zhang, Bowei and Lin, Chaofan and Dong, Chen and others},
  journal={arXiv preprint arXiv:2512.02556},
  year={2025}
}
\bibliographystyle{apalike}

\appendix

\section{The Cost of Running LLMs -- a FLOPs Analysis}\label{app:flops}

This appendix provides a coarse-grained floating point operations (FLOPs)
comparison between a message-passing graph neural network (GNN) and a
middle-sized LLM, in particular \texttt{GPT-OSS-120B} \citep{openai2025gptoss},
under simplified but explicit assumptions. The goal is to illustrate
order-of-magnitude compute differences incurred by using an unnecessarily large
architecture rather than precise hardware-level performance. We note that this
is a smaller GPT-class LLM than the ones we used, so the LLM compute flops are
quite optimistic.  

\subsection{GNN Assumptions and FLOPs calculation}

We consider a $L=6$ layer message-passing GNN with hidden dimension $d=64$
\citep[Appendix C]{galkin2023towards}. Let $N$ denote the number of nodes and
$E$ denote the number of edges in the graph.

Each GNN layer consists of the following operations:
\begin{align*}
    m_{ij} &= h_i + h_j \\
    m_i &= \sum_{j \in \mathcal{N}(i)} m_{ij} \\
    h_i' &= \mathrm{MLP}(m_i)
\end{align*}

The update MLP is a 2-layer perceptron with dimensions $64 \rightarrow 64
\rightarrow 64$, applied independently to each node. As this is strictly a FLOP
analysis, we do not account for irregular access patterns, which can influence
wall-clock GNN runtime.

\paragraph{Message computation}
For each edge, an elementwise addition of two $d$-dimensional vectors is performed:
\[
\mathrm{FLOPs}_{\text{msg}} = E \times d
\]

\paragraph{Aggregation}
Aggregation is implemented as a sum over incoming messages, incurring one addition per feature per edge:
\[
\mathrm{FLOPs}_{\text{aggr}} = E \times d
\]

\paragraph{Update MLP}
Each linear layer of size $d \times d$ requires approximately $2d^2$ FLOPs per node (counting one multiply and one add). For a two-layer MLP:
\[
\mathrm{FLOPs}_{\text{MLP}} \approx 2 \times 2 d^2 N = 4 d^2 N
\]

Combining all terms and substituting for $d=64$ gives:
\[
\mathrm{FLOPs}_{\text{GNN, layer}} \approx 2Ed + 4d^2N = 128E + 16384N
\]

For $L=6$ layers:
\[
\mathrm{FLOPs}_{\text{GNN}} \approx 6(128E + 16384N) = 768E + 98304N
\]

\subsection{LLM Assumptions and FLOPs calculation}

We compare against a \texttt{GPT-OSS-120B}, with the following \emph{overly}
simplifying assumptions:
\begin{itemize}
    \item Active parameters per token are $P_{\text{active}} \approx 5.1 \times
        10^9$ as per \citet[Table 1]{openai2025gptoss}.
    \item FLOPs per token are twice the active parameters -- one add and one
        multiply per active parameter.
        \[
        \mathrm{FLOPs}_{\text{token}} \approx 10.2 \times 10^9
        \]
    \item Input sequence length is $T = N + E$ tokens.
    \item One output token is generated
    \item Quadratic attention costs are ignored, assuming some perfect
        KV-caching/compression algorithm. Attention would otherwise further
        disadvantage the LLM at large scales.
\end{itemize}
These additional assumptions further lower the bound on total LLM compute.

\paragraph{LLM FLOPs}

Total FLOPs for prompt prefill and one decoding step:
\[
    \mathrm{FLOPs}_{\text{LLM}} \approx \mathrm{FLOPs}_{\text{token}}\times(T + 1) \approx 10.2 \times 10^9\times (N + E)
\]

\subsection{Numerical Example}

Assume we have a graph with $N = 3{,}000$ nodes and $E = 30{,}000$ edges. This
is considerably smaller than Wikidata or our top-$30{,}000$ PPR graph.

\begin{align*}
\mathrm{FLOPs}_{\text{GNN}} &= 768 \times 30{,}000 + 98{,}304 \times 3{,}000 \\
&\approx 3.18 \times 10^8 \;\text{FLOPs}\\
\mathrm{FLOPs}_{\text{LLM}} &= 10.2 \times 10^9 \times 33{,}000 \\
&\approx 3.37 \times 10^{14} \;\text{FLOPs}
\end{align*}

The ratio of compute is:
\[
\frac{\mathrm{FLOPs}_{\text{LLM}}}{\mathrm{FLOPs}_{\text{GNN}}}
\approx 1.06 \times 10^6
\]

Thus, for the considered graph size and architectures, a single inference pass
of the LLM requires approximately six orders of magnitude more floating point
operations than the full 6-layer GNN.

\section{Seed Entity Personalised PageRank}\label{app:seppr}

\begin{algorithm}[t]
\caption{Seed Entity Personalised PageRank}
\label{alg:ppr}
\begin{algorithmic}[1]
\REQUIRE Graph $\gG=(\gV,\gE)$, seed set $\gS$: either (i) crisp set $\gS \subseteq \gV$ if ground-truth entities known, or (ii) set of fuzzy sets $\{\{p_i(v)\}_{v \in \gV}\}_i$ from entity linking (see \cref{eq:RBF_norm}), damping factor $\alpha=0.85$, $T=5$, $k$
\ENSURE Top-$k$ nodes sorted by PPR scores
\STATE $\vx_0 \gets \vzero$ \COMMENT{Initialize probability vector}
\IF{$\gS$ is crisp set}
    \STATE $\vx_0[v] \gets 1/|\gS|$ for all $v \in \gS$ \COMMENT{Case (i): uniform weights}
\ELSE
    \STATE $\vx_0[v] \gets \sum_i p_i(v)$ for all $v \in \gV$ \COMMENT{Case (ii): Combine probabilities across fuzzy sets}
    \STATE $\vx_0 \gets \vx_0 / \sum_{u \in \gV} \vx_0[u]$ \COMMENT{Renormalize}
\ENDIF
\STATE $\vx \gets \vx_0$
\FOR{$t = 1$ to $T$}
    \STATE $\vx'[v] \gets \sum_{(u,v) \in \gE} \frac{\vx[u]}{\mathrm{degree}[u]}$ for all $v \in \gV$ \COMMENT{Diffuse}
    \STATE $\vx \gets \alpha \vx' + (1-\alpha)\vx_0$ \COMMENT{Teleport to starting nodes with probability $\alpha$}
\ENDFOR
\STATE \textbf{return} top-$k$ nodes by $\vx$ values
\end{algorithmic}
\end{algorithm}

We present our implementation in \cref{alg:ppr}. 
Differently from past works:

\begin{itemize}
    \item when we use ground-truth seed entities, we initialise the boundary condition $\vx_0$ to be a uniform signal placed on the seed entities;
    \item when entity linking is used instead, $\vx_0[v]$ is set to be the (normalized) sum of
        probabilities of $v$ being one of the seed entities. 
\end{itemize}

After initialisation, $T$ probability diffusion steps take place and then the
knowledge graph is subsampled to the top-$k$ nodes with highest $\vx[v]$.
Differently from \ULTRAQUERYnospace, our PPR algorithm processes only a mono-dimensional signal, and as a result it can fit on a single GPU (GH200 96GB), even for web scale KGs such as Wikidata.

In our experiments, we set $k = 30,000$. However, the number of edges in the subgraphs induced by the PPR nodes cause out-of-memory issues for certain baselines. To remedy that, we prune the number of PPR nodes to ensure that there are no more than 500,000 edges in the subgraphs for the baselines.

\section{Query Language and Validity}
\label{app:query_language_comparison}
\label{app:betae_format}

The query language is part of the LLM--executor interface rather than a standalone claim of novelty. Its role is to make first-order logical forms easy for the LLM to emit and easy for the executor to parse.

\paragraph{BetaE-style nesting.}
\Cref{fig:betae_format} shows the Turing Award example from \cref{fig:vis_abstract} in the BetaE-style format used by the original \ULTRAQUERY implementation. Projections follow the left-hand grammar in \cref{fig:query_language_comparison}. The LLM must emit the full nested expression; the visual breakdown is included only for readability.

\begin{figure}[H]
    \centering
    \begin{tikzpicture}[
    textbox/.style={rectangle, draw=black!30, fill=Gray!10, inner sep=8pt, font=\small, align=left}
]

\node[textbox] (query) at (0, 0) {
    \textbf{Full Query:}\\[0.3em]
    {\ttfamily\small
    \textcolor{YellowOrange!80}{$($}\textcolor{Orange!80}{$($}\textcolor{ForestGreen!70}{$($}\textcolor{SkyBlue}{TA}\textcolor{ForestGreen!70}{$,\ ($}\textcolor{OliveGreen!90}{P1\_inv}\textcolor{ForestGreen!70}{$,))$}\textcolor{Orange!80}{$,\ $}\textcolor{ForestGreen!70}{$($}\textcolor{Cerulean}{DL}\textcolor{ForestGreen!70}{$,\ ($}\textcolor{OliveGreen!90}{P2\_inv}\textcolor{ForestGreen!70}{$,))$}\textcolor{Orange!80}{$)$}\textcolor{YellowOrange!80}{$,\ ($}\textcolor{BrickRed!90}{P4}\textcolor{YellowOrange!80}{$,))$}
    }\\[0.5em]
    \textbf{Breakdown:}\\[0.3em]
    \textcolor{ForestGreen!70}{\textbf{Projection 1:}} {\ttfamily\small\textcolor{ForestGreen!70}{$($}\textcolor{SkyBlue}{TA}\textcolor{ForestGreen!70}{$,\ ($}\textcolor{OliveGreen!90}{P1\_inv}\textcolor{ForestGreen!70}{$,))$}}\\[0.2em]
    \textcolor{ForestGreen!70}{\textbf{Projection 2:}} {\ttfamily\small\textcolor{ForestGreen!70}{$($}\textcolor{SkyBlue}{DL}\textcolor{ForestGreen!70}{$,\ ($}\textcolor{OliveGreen!90}{P2\_inv}\textcolor{ForestGreen!70}{$,))$}}\\[0.2em]
    \textcolor{Orange!70}{\textbf{Intersection:}} {\ttfamily\small\textcolor{Orange!70}{$($}\textit{Proj 1}\textcolor{Orange!70}{$,\ $}\textit{Proj 2}\textcolor{Orange!70}{$)$}}\\[0.2em]
    \textcolor{BrickRed!70}{\textbf{Final Projection:}} {\ttfamily\small\textcolor{BrickRed!70}{$($}\textit{Intersection}\textcolor{BrickRed!70}{$,\ ($}\textcolor{BrickRed!90}{P4}\textcolor{BrickRed!70}{$,))$}}
};

\end{tikzpicture}
    \caption{
        BetaE format representation of the Turing Award query example from
        \cref{fig:vis_abstract}. Entities are abbreviated.
    }\label{fig:betae_format}
\end{figure}

\paragraph{Validity analysis.}
\Cref{tab:dsl_validity_ablation} shows that the BetaE-style format is already reliable on simple path queries, but becomes error-prone on nested classes. The flatter syntax is meant to reduce these parse failures, not to make arbitrary natural-language questions executable. The executor still handles first-order queries; questions requiring counting or genuinely second-order reasoning remain outside the direct query language and must either be approximated by retrieving a relevant set for arbitration or left to future extensions. Hallucinated entities or relation handles are also still counted as errors downstream.

\begin{table}[H]
\centering
\caption{Invalid LLM-generated queries under the BetaE-style format for representative GTSQA classes.}
\label{tab:dsl_validity_ablation}
\begin{tabular}{lcccc}
\toprule
\textbf{Class} & \textbf{$n$} & \textbf{Valid} & \textbf{Non-parseable} & \textbf{Invalid \%} \\
\midrule
{\scriptsize\texttt{(1)}} & 150 & 150 & 0 & 0.0 \\
{\scriptsize\texttt{(2)}} & 150 & 150 & 0 & 0.0 \\
{\scriptsize\texttt{((1)(1))(1)}} & 60 & 48 & 12 & 20.0 \\
{\scriptsize\texttt{(2(1)(1))}} & 55 & 40 & 15 & 27.3 \\
{\scriptsize\texttt{((2)(1))}} & 65 & 55 & 10 & 15.4 \\
\bottomrule
\end{tabular}
\end{table}

\paragraph{Flattened DSL.}
The DSL keeps the same first-order query scope but reduces bracket nesting. The original BetaE-style format uses nested tuples for projections and binary intersections; by contrast, our DSL uses infix notation with \texttt{->} for projections and $n$-ary tuples for intersections. Both formats support inverse projections through an \texttt{\_inv} suffix. In the Turing Award example, the flatter DSL reduces the maximum nesting depth from $4$ to $1$ while preserving the same logical structure.

\begin{figure}[H]
    \centering
    \IfFileExists{tikz-figures/query_language_comparison_iclr.tex}{%
        \begin{tikzpicture}[
    textbox/.style={rectangle, draw=black!30, fill=Gray!10, inner sep=8pt, font=\small, align=left},
    label/.style={font=\footnotesize\bfseries, text=black}
]

\node[textbox, text width=.95\textwidth] (combined) at (0, 0) {
    \begin{minipage}[t]{0.48\textwidth}
        \textbf{Query} ::= \textit{Projection} $|$ \textit{Intersection}\\[0.4em]
        
        \textbf{Projection} ::= $($ \textit{Entity}, $($ \textit{Relation}, $))$ \quad (leaf)\\[0.3em]
        \hspace{1em} $|$ $($ \textit{Query}, $($ \textit{Relation}, $))$ \quad (chained)\\[0.4em]
        
        \textbf{Intersection} ::= $($ \textit{Query}, \textit{Query} $)$\\[0.4em]
        
        \textbf{Entity} ::= \texttt{"Q<digits>"}\\[0.3em]
        \textbf{Relation} ::= \texttt{"P<digits>"} $|$ \texttt{"P<digits>\_inv"}
    \end{minipage}
    \hfill
    \vrule
    \hfill
    \begin{minipage}[t]{0.48\textwidth}
        \textbf{Query} ::= \textit{Projection} $|$ \textit{Intersection}\\[0.4em]
        
        \textbf{Projection} ::= \textit{Entity} \texttt{->} \textit{Relation} \quad (leaf)\\[0.3em]
        \hspace{1em} $|$ \textit{Query} \texttt{->} \textit{Relation} \quad (chained)\\[0.4em]
        
        \textbf{Intersection} ::= \texttt{AND(} \textit{Query} \texttt{,} \textit{Query} \texttt{[,} \textit{Query} \texttt{...]} \texttt{)}\\[0.4em]
        
        \textbf{Entity} ::= \texttt{"Q<digits>"}\\[0.3em]
        \textbf{Relation} ::= \texttt{"P<digits>"} $|$ \texttt{"P<digits>\_inv"}
    \end{minipage}\\[0.4em]
    \rule{\textwidth}{0.4pt}\\[0.3em]
    \begin{minipage}[t]{0.48\textwidth}
        {\scriptsize\ttfamily(((TA, (P1\_inv,)), (DL, (P2\_inv,))), (P4,))}
    \end{minipage}
    \hfill
    \vrule
    \hfill
    \begin{minipage}[t]{0.48\textwidth}
        {\scriptsize\ttfamily AND(TA -> P1\_inv, DL -> P2\_inv) -> P4}
    \end{minipage}
};


\end{tikzpicture}%
    }{%
        \begin{tikzpicture}[
    textbox/.style={rectangle, draw=black!30, fill=Gray!10, inner sep=8pt, font=\small, align=left},
    label/.style={font=\footnotesize\bfseries, text=black}
]

\node[textbox, text width=.95\textwidth] (combined) at (0, 0) {
    \begin{minipage}[t]{0.48\textwidth}
        \textbf{Query} ::= \textit{Projection} $|$ \textit{Intersection}\\[0.4em]
        
        \textbf{Projection} ::= $($ \textit{Entity}, $($ \textit{Relation}, $))$ \quad (leaf)\\[0.3em]
        \hspace{1em} $|$ $($ \textit{Query}, $($ \textit{Relation}, $))$ \quad (chained)\\[0.4em]
        
        \textbf{Intersection} ::= $($ \textit{Query}, \textit{Query} $)$\\[0.4em]
        
        \textbf{Entity} ::= \texttt{"Q<digits>"}\\[0.3em]
        \textbf{Relation} ::= \texttt{"P<digits>"} $|$ \texttt{"P<digits>\_inv"}
    \end{minipage}
    \hfill
    \vrule
    \hfill
    \begin{minipage}[t]{0.48\textwidth}
        \textbf{Query} ::= \textit{Projection} $|$ \textit{Intersection}\\[0.4em]
        
        \textbf{Projection} ::= \textit{Entity} \texttt{->} \textit{Relation} \quad (leaf)\\[0.3em]
        \hspace{1em} $|$ \textit{Query} \texttt{->} \textit{Relation} \quad (chained)\\[0.4em]
        
        \textbf{Intersection} ::= \texttt{AND(} \textit{Query} \texttt{,} \textit{Query} \texttt{[,} \textit{Query} \texttt{...]} \texttt{)}\\[0.4em]
        
        \textbf{Entity} ::= \texttt{"Q<digits>"}\\[0.3em]
        \textbf{Relation} ::= \texttt{"P<digits>"} $|$ \texttt{"P<digits>\_inv"}
    \end{minipage}\\[0.4em]
    \rule{\textwidth}{0.4pt}\\[0.3em]
    \begin{minipage}[t]{0.48\textwidth}
        {\scriptsize\ttfamily(((TA, (P1\_inv,)), (DL, (P2\_inv,))), (P4,))}
    \end{minipage}
    \hfill
    \vrule
    \hfill
    \begin{minipage}[t]{0.48\textwidth}
        {\scriptsize\ttfamily AND(TA -> P1\_inv, DL -> P2\_inv) -> P4}
    \end{minipage}
};


\end{tikzpicture}%
    }
    \caption{
        Haskell-like grammar definitions for the old BetaE format (left) and our preferred DSL
        (right), with the transformed Turing Award example below.
    }\label{fig:query_language_comparison}
\end{figure}

In the same targeted analysis, prompting GPT-5 with the \ULTRAG DSL resulted in only two erroneous queries across all classes. Thus, the DSL does not eliminate all query-construction inaccuracies, but it makes syntactic invalidity a small residual source of error instead of the $15$--$30\%$ failure mode observed with the nested tuple format.

\section{Neural vs Symbolic}\label{app:wikidata_neural_symbolic}

\begin{table}[H]
\centering
\caption{Comparison of neural (UltraQuery) vs symbolic query execution on
LLM-generated queries from GTSQA using WikiKG2 subgraphs (in \%). Class notation as in
\citeauthor{cattaneo2025ground} -- number inside brackets denotes hops\Kheeran{projections},
concatenation of expression denotes intersection. Both executors receive
identical queries generated by the LLM.}
\label{tab:neural_vs_symbolic}
\resizebox{\textwidth}{!}{
\begin{tabular}{lccccccccccccccccccc}
\toprule
\multirow{2}{*}{\textbf{Metric}} & \multicolumn{5}{c}{$\#$ hops = 1} & \multicolumn{6}{c}{$\#$ hops = 2} & \multicolumn{5}{c}{$\#$ hops = 3} & \multicolumn{2}{c}{$\#$ hops = 4} \\
\cmidrule(lr){2-6} \cmidrule(lr){7-12} \cmidrule(lr){13-17} \cmidrule(lr){18-19}
 & {\scriptsize\texttt{(1)}} & {\scriptsize\texttt{(1)(1)}} & {\scriptsize\texttt{(1)(1)(1)}} & {\scriptsize\texttt{(1)(1)(1)(1)}} & {\scriptsize Avg} & {\scriptsize\texttt{(2)}} & {\scriptsize\texttt{(2)(1)}} & {\scriptsize\texttt{((1)(1))}} & {\scriptsize\texttt{(2)(2)}} & {\scriptsize\texttt{((1)(1))(1)}} & {\scriptsize Avg} & {\scriptsize\texttt{(3)}} & {\scriptsize\texttt{((2)(1))}} & {\scriptsize\texttt{(3)(1)}} & {\scriptsize\texttt{(2(1)(1))}} & {\scriptsize Avg} & {\scriptsize\texttt{(4)}} & {\scriptsize Avg} \\
\midrule
\multicolumn{19}{c}{\textbf{Neural executor (UltraQuery)}} \\
\midrule
MRR & 93.25 & 94.48 & 97.96 & 96.36 & 95.51 & 85.24 & 86.19 & 93.28 & 95.17 & 84.01 & 88.78 & 82.78 & 92.00 & 90.30 & 88.35 & 88.36 & 82.54 & 82.54 \\
Hit@1 & 91.33 & 93.17 & 96.94 & 95.93 & 94.34 & 82.67 & 82.48 & 92.00 & 93.05 & 80.00 & 86.04 & 77.50 & 87.69 & 88.00 & 85.00 & 84.55 & 76.22 & 76.22 \\
Hit@3 & 94.67 & 95.00 & 98.98 & 96.75 & 96.35 & 87.33 & 87.99 & 93.60 & 97.60 & 86.17 & 90.54 & 87.87 & 93.85 & 90.67 & 92.27 & 91.16 & 88.22 & 88.22 \\
Hit@10 & 96.00 & 97.78 & 98.98 & 96.75 & 97.38 & 88.67 & 92.13 & 96.80 & 97.84 & 93.33 & 93.75 & 90.65 & 100.00 & 96.00 & 93.64 & 95.07 & 92.22 & 92.22 \\
\midrule
\multicolumn{19}{c}{\textbf{Symbolic executor}} \\
\midrule
MRR & 90.03 & 79.05 & 80.13 & 82.12 & 82.83 & 75.11 & 59.51 & 77.62 & 74.84 & 64.96 & 70.41 & 56.13 & 76.95 & 58.78 & 62.28 & 63.53 & 64.08 & 64.08 \\
Hit@1 & 90.00 & 78.67 & 79.59 & 82.11 & 82.59 & 74.67 & 59.06 & 77.60 & 74.82 & 63.33 & 69.90 & 55.65 & 76.92 & 58.00 & 61.67 & 63.06 & 64.00 & 64.00 \\
Hit@3 & 90.00 & 79.33 & 80.61 & 82.11 & 83.01 & 75.33 & 59.84 & 77.60 & 74.82 & 66.67 & 70.85 & 56.20 & 76.92 & 59.33 & 63.03 & 63.87 & 64.00 & 64.00 \\
Hit@10 & 90.00 & 79.33 & 80.61 & 82.11 & 83.01 & 76.00 & 60.24 & 77.60 & 74.82 & 66.67 & 71.07 & 56.76 & 76.92 & 59.33 & 63.03 & 64.01 & 64.00 & 64.00 \\
\midrule
\multicolumn{19}{c}{\textbf{Improvement (Neural - Symbolic)}} \\
\midrule
MRR & +3.22 & +15.43 & +17.83 & +14.24 & +12.68 & +10.13 & +26.68 & +15.66 & +20.33 & +19.05 & +18.37 & +26.65 & +15.05 & +31.52 & +26.07 & +24.82 & +18.46 & +18.46 \\
Hit@1 & +1.33 & +14.50 & +17.35 & +13.82 & +11.75 & +8.00 & +23.42 & +14.40 & +18.23 & +16.67 & +16.14 & +21.85 & +10.77 & +30.00 & +23.33 & +21.49 & +12.22 & +12.22 \\
Hit@3 & +4.67 & +15.67 & +18.37 & +14.64 & +13.34 & +12.00 & +28.15 & +16.00 & +22.78 & +19.50 & +19.69 & +31.67 & +16.93 & +31.34 & +29.24 & +27.29 & +24.22 & +24.22 \\
Hit@10 & +6.00 & +18.45 & +18.37 & +14.64 & +14.37 & +12.67 & +31.89 & +19.20 & +23.02 & +26.66 & +22.69 & +33.89 & +23.08 & +36.67 & +30.61 & +31.06 & +28.22 & +28.22 \\
\bottomrule
\end{tabular}
}
\end{table}

\begin{table}[H]
\centering
\caption{Comparison of neural (UltraQuery) vs symbolic query execution on LLM-generated queries from GTSQA on PPR subgraphs extracted from Wikidata (in \%). Both executors receive identical queries generated by the LLM.}
\label{tab:wikidata_neural_vs_symbolic}
\resizebox{\textwidth}{!}{
\begin{tabular}{lccccccccccccccccccc}
\toprule
\multirow{2}{*}{\textbf{Metric}} & \multicolumn{5}{c}{$\#$ hops = 1} & \multicolumn{6}{c}{$\#$ hops = 2} & \multicolumn{5}{c}{$\#$ hops = 3} & \multicolumn{2}{c}{$\#$ hops = 4} \\
\cmidrule(lr){2-6} \cmidrule(lr){7-12} \cmidrule(lr){13-17} \cmidrule(lr){18-19}
 & {\scriptsize\texttt{(1)}} & {\scriptsize\texttt{(1)(1)}} & {\scriptsize\texttt{(1)(1)(1)}} & {\scriptsize\texttt{(1)(1)(1)(1)}} & {\scriptsize Avg} & {\scriptsize\texttt{(2)}} & {\scriptsize\texttt{(2)(1)}} & {\scriptsize\texttt{((1)(1))}} & {\scriptsize\texttt{(2)(2)}} & {\scriptsize\texttt{((1)(1))(1)}} & {\scriptsize Avg} & {\scriptsize\texttt{(3)}} & {\scriptsize\texttt{((2)(1))}} & {\scriptsize\texttt{(3)(1)}} & {\scriptsize\texttt{(2(1)(1))}} & {\scriptsize Avg} & {\scriptsize\texttt{(4)}} & {\scriptsize Avg} \\
\midrule
\multicolumn{19}{c}{\textbf{Neural executor (UltraQuery)}} \\
\midrule
MRR & 91.07 & 86.74 & 91.14 & 90.19 & 89.78 & 77.01 & 78.43 & 88.19 & 77.87 & 74.89 & 79.28 & 61.04 & 75.81 & 75.44 & 72.64 & 71.23 & 70.69 & 70.69 \\
Hit@1 & 90.00 & 83.65 & 88.61 & 87.26 & 87.38 & 74.00 & 74.69 & 85.60 & 69.54 & 69.17 & 74.60 & 54.44 & 66.15 & 66.67 & 64.97 & 63.06 & 65.22 & 65.22 \\
Hit@3 & 91.33 & 88.38 & 93.88 & 93.09 & 91.67 & 78.67 & 81.73 & 89.60 & 82.73 & 78.61 & 82.27 & 65.28 & 86.15 & 81.33 & 76.50 & 77.31 & 73.44 & 73.44 \\
Hit@10 & 93.33 & 92.38 & 93.88 & 94.31 & 93.47 & 85.33 & 84.88 & 94.40 & 93.17 & 83.00 & 88.16 & 72.22 & 89.23 & 90.00 & 86.32 & 84.44 & 82.44 & 82.44 \\
\midrule
\multicolumn{19}{c}{\textbf{Symbolic executor}} \\
\midrule
MRR & 84.45 & 74.30 & 71.01 & 77.89 & 76.91 & 70.05 & 58.12 & 76.27 & 73.94 & 62.94 & 68.26 & 45.96 & 63.97 & 57.84 & 64.29 & 58.02 & 65.04 & 65.04 \\
Hit@1 & 84.00 & 73.33 & 69.73 & 77.24 & 76.08 & 69.33 & 57.32 & 76.00 & 72.66 & 61.67 & 67.40 & 44.17 & 63.08 & 56.00 & 62.68 & 56.48 & 64.00 & 64.00 \\
Hit@3 & 84.67 & 74.67 & 72.45 & 78.59 & 77.59 & 70.67 & 58.90 & 76.80 & 75.54 & 63.33 & 69.05 & 47.50 & 64.62 & 59.33 & 67.10 & 59.64 & 64.67 & 64.67 \\
Hit@10 & 85.33 & 76.00 & 72.45 & 78.86 & 78.16 & 72.00 & 59.69 & 76.80 & 76.26 & 65.00 & 69.95 & 48.61 & 64.62 & 60.00 & 67.10 & 60.08 & 67.78 & 67.78 \\
\midrule
\multicolumn{19}{c}{\textbf{Improvement (Neural - Symbolic)}} \\
\midrule
MRR & +6.62 & +12.44 & +20.13 & +12.30 & +12.87 & +6.96 & +20.31 & +11.92 & +3.93 & +11.95 & +11.01 & +15.08 & +11.84 & +17.60 & +8.35 & +13.22 & +5.65 & +5.65 \\
Hit@1 & +6.00 & +10.32 & +18.88 & +10.02 & +11.30 & +4.67 & +17.37 & +9.60 & -3.12 & +7.50 & +7.20 & +10.27 & +3.07 & +10.67 & +2.29 & +6.57 & +1.22 & +1.22 \\
Hit@3 & +6.66 & +13.71 & +21.43 & +14.50 & +14.07 & +8.00 & +22.83 & +12.80 & +7.19 & +15.28 & +13.22 & +17.78 & +21.53 & +22.00 & +9.40 & +17.68 & +8.77 & +8.77 \\
Hit@10 & +8.00 & +16.38 & +21.43 & +15.45 & +15.32 & +13.33 & +25.19 & +17.60 & +16.91 & +18.00 & +18.21 & +23.61 & +24.61 & +30.00 & +19.22 & +24.36 & +14.66 & +14.66 \\
\bottomrule
\end{tabular}
}
\end{table}

As discussed in the main text, we present the full table of results for
WikiKG2. The neural executor provides a clear advantage over symbolic
execution, with average gains (computed by averaging the per-hop “Avg” columns
in the tables) of $18.58\%$ (MRR), $15.40\%$ (Hit@1), $21.14\%$ (Hit@3), and
$24.09\%$ (Hit@10).

In addition to full results on WikiKG2 in \cref{tab:neural_vs_symbolic}, in \cref{tab:wikidata_neural_vs_symbolic} we provide the results of neural vs symbolic query execution performance on PPR subgraphs extracted from Wikidata using seed entities available in \cite{cattaneo2025ground}. As already highlighted in the main text of the paper, the neural query executor consistently outperforms symbolic execution across the considered metrics and classes of queries (only exception on Wikidata being Hit@1 for class (2)(2), where the  symbolic executor has a gain of $3.12\%$). 
Compared to WikiKG2 subgraphs, the absolute performance is generally lower on Wikidata, which can be the result of both an increased amount of noise in the queries (as a result of the larger set of entities and relation types available in Wikidata) and possibly noise in the extracted subgraph due to the additional PPR extraction phase. This said, while performance degrade to some extent, they still appear good even in this more challenging scenario, and the relative advantage of neural over symbolic execution remains substantial.

The same conclusion holds at the final-answer level. To isolate the executor contribution after downstream arbitration, we keep the rest of the \ULTRAGnospace-OTS pipeline fixed and replace \ULTRAQUERY with symbolic execution in the GTSQA (Wikidata) ground-truth seed setting. \Cref{tab:symbolic_executor_replacement} shows that this replacement decreases Hits/Recall/F1 by 7.58/6.41/5.16 points.

\begin{table}[H]
\centering
\caption{Executor replacement ablation on GTSQA (Wikidata) in the ground-truth seed setting. Scores are final-answer Hits/Recall/F1, so the comparison includes the downstream arbitration stage.}
\label{tab:symbolic_executor_replacement}
\begin{tabular}{lccc}
\toprule
\textbf{Executor} & \textbf{Hits} & \textbf{Recall} & \textbf{F1} \\
\midrule
\ULTRAQUERY & 86.74 & 84.47 & 82.08 \\
Symbolic execution & 79.16 & 78.06 & 76.92 \\
\midrule
Drop & -7.58 & -6.41 & -5.16 \\
\bottomrule
\end{tabular}
\end{table}



\section{Ground-truth queries vs LLM generated queries on WikiKG2}
\label{app:neural_vs_symbolic}

\begin{table}[H]
\centering
\caption{\ULTRAQUERY per-class performance metrics on WikiKG2 subgraphs contained in GTSQA (in \%). To generate ground-truth structured queries (i.e. queries provided by the oracle), given a subgraph with the answer as the root node, we perform a breadth-first traversal, identify leaves, and work upwards to construct the query.}
\label{tab:query_type_results}
\resizebox{\textwidth}{!}{
\begin{tabular}{lccccccccccccccccccc}
\toprule
\multirow{2}{*}{\textbf{Metric}} & \multicolumn{5}{c}{$\#$ hops = 1} & \multicolumn{6}{c}{$\#$ hops = 2} & \multicolumn{5}{c}{$\#$ hops = 3} & \multicolumn{2}{c}{$\#$ hops = 4} \\
\cmidrule(lr){2-6} \cmidrule(lr){7-12} \cmidrule(lr){13-17} \cmidrule(lr){18-19}
 & {\scriptsize\texttt{(1)}} & {\scriptsize\texttt{(1)(1)}} & {\scriptsize\texttt{(1)(1)(1)}} & {\scriptsize\texttt{(1)(1)(1)(1)}} & {\scriptsize Avg} & {\scriptsize\texttt{(2)}} & {\scriptsize\texttt{(2)(1)}} & {\scriptsize\texttt{((1)(1))}} & {\scriptsize\texttt{(2)(2)}} & {\scriptsize\texttt{((1)(1))(1)}} & {\scriptsize Avg} & {\scriptsize\texttt{(3)}} & {\scriptsize\texttt{((2)(1))}} & {\scriptsize\texttt{(3)(1)}} & {\scriptsize\texttt{(2(1)(1))}} & {\scriptsize Avg} & {\scriptsize\texttt{(4)}} & {\scriptsize Avg} \\
\midrule
\multicolumn{19}{c}{\textbf{With ground-truth queries}} \\
\midrule
MRR & 99.33 & 100.00 & 100.00 & 100.00 & 99.83 & 99.56 & 100.00 & 100.00 & 100.00 & 100.00 & 99.91 & 98.61 & 100.00 & 99.83 & 97.12 & 98.89 & 96.07 & 96.07 \\
Hit@1 & 98.67 & 100.00 & 100.00 & 100.00 & 99.67 & 99.33 & 100.00 & 100.00 & 100.00 & 100.00 & 99.87 & 97.78 & 100.00 & 99.67 & 94.55 & 98.00 & 94.00 & 94.00 \\
Hit@3 & 100.00 & 100.00 & 100.00 & 100.00 & 100.00 & 100.00 & 100.00 & 100.00 & 100.00 & 100.00 & 100.00 & 99.44 & 100.00 & 100.00 & 100.00 & 99.86 & 98.00 & 98.00 \\
Hit@10 & 100.00 & 100.00 & 100.00 & 100.00 & 100.00 & 100.00 & 100.00 & 100.00 & 100.00 & 100.00 & 100.00 & 99.44 & 100.00 & 100.00 & 100.00 & 99.86 & 98.00 & 98.00 \\
\midrule
\multicolumn{19}{c}{\textbf{With LLM-generated queries}} \\
\midrule
MRR & 93.25 & 94.48 & 97.96 & 96.36 & 95.51 & 85.24 & 86.19 & 93.28 & 95.17 & 84.01 & 88.78 & 82.78 & 92.00 & 90.30 & 88.35 & 88.36 & 82.54 & 82.54 \\
Hit@1 & 91.33 & 93.17 & 96.94 & 95.93 & 94.34 & 82.67 & 82.48 & 92.00 & 93.05 & 80.00 & 86.04 & 77.50 & 87.69 & 88.00 & 85.00 & 84.55 & 76.22 & 76.22 \\
Hit@3 & 94.67 & 95.00 & 98.98 & 96.75 & 96.35 & 87.33 & 87.99 & 93.60 & 97.60 & 86.17 & 90.54 & 87.87 & 93.85 & 90.67 & 92.27 & 91.16 & 88.22 & 88.22 \\
Hit@10 & 96.00 & 97.78 & 98.98 & 96.75 & 97.38 & 88.67 & 92.13 & 96.80 & 97.84 & 93.33 & 93.75 & 90.65 & 100.00 & 96.00 & 93.64 & 95.07 & 92.22 & 92.22 \\
\midrule
\multicolumn{19}{c}{\textbf{Improvement (LLM generated queries - GT queries)}} \\
\midrule
MRR & -6.08 & -5.52 & -2.04 & -3.64 & -4.32 & -14.32 & -13.81 & -6.72 & -4.83 & -15.99 & -11.13 & -15.83 & -8.00 & -9.53 & -8.77 & -10.53 & -13.53 & -13.53 \\
Hit@1 & -7.34 & -6.83 & -3.06 & -4.07 & -5.33 & -16.66 & -17.52 & -8.00 & -6.95 & -20.00 & -13.83 & -20.28 & -12.31 & -11.67 & -9.55 & -13.45 & -17.78 & -17.78 \\
Hit@3 & -5.33 & -5.00 & -1.02 & -3.25 & -3.65 & -12.67 & -12.01 & -6.40 & -2.40 & -13.83 & -9.46 & -11.57 & -6.15 & -9.33 & -7.73 & -8.69 & -9.78 & -9.78 \\
Hit@10 & -4.00 & -2.22 & -1.02 & -3.25 & -2.62 & -11.33 & -7.87 & -3.20 & -2.16 & -6.67 & -6.25 & -8.79 & 0.00 & -4.00 & -6.36 & -4.79 & -5.78 & -5.78 \\
\bottomrule
\end{tabular}
}
\end{table}


In \cref{tab:query_type_results} we compare performance of our chosen neural query executor (\ULTRAQUERYnospace) when processing GT queries vs LLM-generated queries on the WikiKG2 subgraphs of GTSQA \cite{cattaneo2025ground}. As we can see, 
the hop-averaged performance gap ("Avg" columns in the table) between ground-truth and LLM-generated queries appears to be generally growing with query complexity, starting at $4.32\%$ of MRR for 1-hop subgraphs, and ending with $13.53\%$ MRR for 4-hops (similar trends also appear for the other considered metrics). This highlights the challenge of query generation for complex multi-hop reasoning. 

\section{ULTRAG with Different LLMs}
\label{app:ultrag_llm_ablation}

In \cref{tab:ultrag_llm_ablation} we provide a comparison of \ULTRAGnospace-OTS performance with different LLMs. For this analysis, we used GTSQA using PPR subgraphs extracted from Wikidata with ground-truth seed entities.


\begin{table}[H]
\centering
\caption{Comparison of \ULTRAGnospace-OTS with different LLMs on GTSQA (Wikidata) using ground-truth seed entities (in \%). PPR graphs have up to 30,000 nodes. With $A \to B$, we describe a setting where we used LLM $A$ for query generation and $B$ for arbitration.}
\label{tab:ultrag_llm_ablation}
\begin{tabular}{lcccc}
\toprule
\textbf{LLM} & \textbf{Hits} & \textbf{Recall} & \textbf{F1} & \textbf{API cost} \\
\midrule
\ULTRAG &       &       &      &  \\
- GPT-5$\times2$ & 86.74 & 84.47 & 82.08 & 0.017\$\\
- GPT-5$\to$GPT-5-mini & 82.37 & 77.83 & 78.41 & 0.009\$\\
- GPT-5-mini$\to$GPT-5 & 80.71 & 78.19 & 75.72 & 0.011\$\\
- DeepSeek-reasoner$\times2$ & 75.77 & 72.40 & 72.05 &  0.005\$\\
- GPT-5-mini$\times2$ & 75.71 & 71.08 & 71.60 & 0.004\$ \\
\bottomrule
\end{tabular}
\end{table}


\section{Runtime costs on WikiKG2 and Wikidata}
\label{app:runtime_costs}

In \cref{tab:efficiency_comparison_KGQAGen} we show average API costs per query, average non-API times, and average number of input / output tokens processed / generated by the LLM (GPT-5) on KGQAGen subgraphs extracted from Wikidata. On this dataset, \ULTRAGnospace-OTS is 15x faster than RoG, 
122x faster than GNN-RAG, and 130x faster than SubgraphRAG, in terms of non-API time per query. In terms of number of input tokens, \ULTRAGnospace-OTS processes between 22x and 50x more tokens compared to the baselines (again due to the presence of relation types in the input prompt and two API calls). However, as it was the case for GTSQA, $96\%$ of these tokens are cached by GPT-5. Interestingly, average number of output tokens and API costs are  $\sim$2x larger for \ULTRAGnospace-OTS compared to the baselines, which appear more efficient on this dataset. We note however, that state-of-the-art performance (Hits$ = 86.07 \%$, Recall$ = 84.45 \%$, F1$= 83.91 \%$) can be still achieved on KGQAGen with \ULTRAGnospace-OTS, at a fraction of the cost (\$0.0036), by using GPT-5-mini for both generation and arbitration in place of GPT-5 (\ULTRAGnospace-OTS implemented with GPT-5-mini is actually the cheapest solution in our comparison). This highlights that, while KGQAGen questions requires larger amounts of output tokens for producing valuable results with our methodology, questions there appear to be simpler: \ULTRAGnospace-OTS with lighter and less skillful LLMs can still outperform baselines that use GPT-5. As it was the case for GTSQA, we also highlight that in an end-to-end pipeline, no additional cost would be required for mention extraction with \ULTRAGnospace-OTS, while all the considered baselines would incur extra costs for this additional step\footnote{If GPT-5 was used for this, all baselines would incur an additional cost of 0.004\$ per query, while also showing an increase of 780 in the average number of output tokens produced.}.

\begin{table}[H]
\caption{Average efficiency per query on PPR subgraphs extracted for KGQAGen-10K from Wikidata with ground-truth seed nodes. All models use GPT-5 as reasoning LLM. Non-API running time measured on GH200 chips, excluding API calls and PPR computation.} 
\label{tab:efficiency_comparison_KGQAGen}
\centering
\begin{tabular}{lccccc}
\toprule
\multirowcell{2}{\textbf{KGQAGen-10K}} & \multirowcell{2}{\textbf{API cost}} & \textbf{\multirowcell{2}{Non-API \\ time (s)}} & \textbf{\multirowcell{2}{Input \\ tokens}} & \textbf{\multirowcell{2}{Output \\ tokens}} & \textbf{\multirowcell{2}{Cache hit \\ \%}} \\
& & & & & \\
\midrule
    RoG  & 0.008\$ & $3.1 \pm 0.9$ & 1.4K & 1.4K & 0 \\
    GNN-RAG  & 0.008\$ & $25.8 \pm 4.4$ & 1.4K & 1.4K & 0\\
    SubgraphRAG (200)  & 0.009\$ & $27.5 \pm 3.9$ & 3.1K & 1.5K & 0 \\
    \textbf{\ULTRAGnospace-OTS} & 0.019\$ & $0.21 \pm 0.1$ & 69K & 2.8K & 96.06\\
    \,\,\,\,\,\,- GPT-5-mini & 0.004\$ & $0.15 \pm 0.1$ & 69K & 2.4K & 95.53\\ 
\bottomrule
\end{tabular}
\end{table}


\section{Hyperparameter Sensitivity}
\label{app:hyperparameter_sensitivity}

\begin{table}[H]
\centering
\caption{Sensitivity of \ULTRAGnospace-OTS to PPR subgraph size and the number of executor candidates passed to the arbitrator. We report F1 on GTSQA (Wikidata) with ground-truth seed entities.}
\label{tab:hyperparameter_sensitivity}
\begin{tabular}{ccc}
\toprule
\textbf{PPR nodes} & \textbf{Candidate top-$k$} & \textbf{F1} \\
\midrule
5K & 10 & 80.99 \\
5K & 50 & 80.33 \\
5K & 100 & 80.82 \\
30K & 10 & 81.45 \\
30K & 50 & \textbf{82.08} \\
30K & 100 & 81.81 \\
50K & 10 & 81.61 \\
50K & 50 & 81.49 \\
50K & 100 & 81.43 \\
\bottomrule
\end{tabular}
\end{table}

The reference setting used in the main experiments is $30$K PPR nodes and top-$50$ executor candidates. Across this grid, the largest F1 change is below two points, suggesting that the reported gains are not a consequence of a narrow hyperparameter choice.

\section{Multi-step Ablation}
\label{app:multistep_ablation}

\begin{table}[H]
\centering
\caption{Single-iteration and multi-iteration \ULTRAGnospace-OTS F1 on GTSQA query classes. Class notation follows \citet{cattaneo2025ground}.}
\label{tab:multistep_ablation}
\resizebox{\textwidth}{!}{
\begin{tabular}{lcccccccccccccc}
\toprule
\textbf{Setting} &
{\scriptsize\texttt{((1)(1))}} &
{\scriptsize\texttt{((1)(1))(1)}} &
{\scriptsize\texttt{((2)(1))}} &
{\scriptsize\texttt{(1)}} &
{\scriptsize\texttt{(1)(1)}} &
{\scriptsize\texttt{(1)(1)(1)}} &
{\scriptsize\texttt{(1)(1)(1)(1)}} &
{\scriptsize\texttt{(2(1)(1))}} &
{\scriptsize\texttt{(2)}} &
{\scriptsize\texttt{(2)(1)}} &
{\scriptsize\texttt{(2)(2)}} &
{\scriptsize\texttt{(3)}} &
{\scriptsize\texttt{(3)(1)}} &
{\scriptsize\texttt{(4)}} \\
\midrule
Single-iter F1 & 93.87 & 77.06 & 88.72 & 96.22 & 90.59 & 89.52 & 84.25 & 89.24 & 93.11 & 84.39 & 94.48 & 85.09 & 88.79 & 83.73 \\
Multi-iter F1 & 93.96 & 82.44 & 92.82 & 96.93 & 91.86 & 91.63 & 85.79 & 85.57 & 94.67 & 87.42 & 95.91 & 87.59 & 92.72 & 91.13 \\
\bottomrule
\end{tabular}
}
\end{table}

The multi-step variant improves overall F1 by about two points, with larger gains on several complex classes. We use the single-iteration version in the main experiments because it is simpler and already sufficient for state-of-the-art performance, but the iterative formulation is empirically useful when additional compute is acceptable.

\section{Arbitrator Ablation}
\label{app:arbitrator_ablation}

\ULTRAQUERY returns ranked probabilities rather than a discrete answer set, so removing the LLM arbitrator requires a hand-designed conversion rule. In \cref{tab:arbitrator_threshold_ablation}, \emph{Top-1} and \emph{Top-5} return the one or five highest-scoring entities, respectively. \emph{Top-$p$} with threshold $\tau$ returns every entity whose executor probability satisfies $p>\tau$; we test $\tau \in \{0.3,0.5,0.7\}$. The \emph{relative threshold} returns all entities whose executor probability is larger than half of the maximum probability, i.e. $p > 0.5p_{\max}$. \emph{LLM arbitrator} is the full \ULTRAGnospace-OTS setting, where GPT-5 receives the ranked candidates and probabilities and produces the final answer set.
\Cref{tab:arbitrator_threshold_ablation} shows that such rules are consistently weaker than LLM arbitration.

\begin{table}[H]
\centering
\caption{F1 of \ULTRAGnospace-OTS when the LLM arbitrator is replaced by deterministic conversions from executor probabilities to answer sets. The full LLM arbitrator remains the most reliable conversion rule.}
\label{tab:arbitrator_threshold_ablation}
\begin{tabular}{lcc}
\toprule
\textbf{Setting} & \textbf{GTSQA} & \textbf{KGQAGen-10k} \\
\midrule
Top-1 & 84.9 & 80.9 \\
Top-5 & 35.3 & 34.0 \\
Top-$p$ 0.3 & 68.6 & 59.6 \\
Top-$p$ 0.5 & 73.4 & 57.0 \\
Top-$p$ 0.7 & 72.1 & 54.7 \\
Relative threshold & 81.1 & 80.8 \\
LLM arbitrator & \textbf{89.3} & \textbf{88.8} \\
\bottomrule
\end{tabular}
\end{table}

The LLM arbitrator is not simply answering from parametric memory: across datasets, at least $94\%$ of generated responses select one of the top-$50$ executor candidates as the final answer.

\section{Randomized Relation Identifiers}
\label{app:relation_id_ablation}

This ablation tests whether relation identifiers act only as parser-facing handles or whether they accidentally provide semantic information. This matters because Wikidata PIDs are stable public identifiers, and an LLM may have memorized associations for common relations. If performance depended on those specific IDs, the interface would be less realistic and the results could partially reflect hidden relation-name knowledge rather than relation selection from the provided labels and descriptions. Randomizing the identifiers per sample removes any such memorized PID semantics while preserving the same query syntax and one-to-one relation lookup interface.

\begin{table}[H]
\centering
\caption{\ULTRAGnospace-OTS performance on GTSQA when relation identifiers are randomly reassigned per sample. Scores are reported as Hits/Recall/F1.}
\label{tab:relation_id_ablation}
\begin{tabular}{lccc}
\toprule
\textbf{Setting} & \textbf{Hits} & \textbf{Recall} & \textbf{F1} \\
\midrule
No randomization & 92.66 & 91.05 & 89.29 \\
Randomized relation IDs & 93.90 & 92.37 & 91.46 \\
\bottomrule
\end{tabular}
\end{table}

For each sample, relation identifiers were randomly sampled without replacement from $[0,9999]$. Performance remains in line with the unshuffled setting, supporting the interpretation that identifiers are only stable handles for query parsing rather than semantic cues or hidden supervision. Samples with hallucinated identifiers are marked erroneous.

\section{WebQSP and CWQ Results}
\label{app:webqsp_cwq_results}

\begin{table}[H]
\centering
\caption{WebQSP results. Scores are in \%; $\Delta$ is relative to the best non-\ULTRAG baseline in each column.}
\label{tab:webqsp_results}
\begin{tabular}{lcccc}
\toprule
\textbf{Model} & \textbf{Hits} & \textbf{Recall} & \textbf{Precision} & \textbf{F1} \\
\midrule
GPT-5 & 73.0 & 53.6 & 66.3 & 55.5 \\
RoG & 88.5 & 78.4 & 84.7 & 78.5 \\
GNN-RAG & 88.8 & 78.1 & 85.4 & 78.7 \\
SubgraphRAG & 87.2 & 76.1 & 82.7 & 76.1 \\
GCR & \textbf{90.0} & 71.0 & 85.4 & 73.0 \\
\ULTRAGnospace-OTS & 88.0 & 77.7 & 82.3 & 76.7 \\
$\Delta$ & -2.0 & -0.7 & -3.1 & -2.0 \\
\bottomrule
\end{tabular}
\end{table}

\begin{table}[H]
\centering
\caption{CWQ results. Scores are in \%; $\Delta$ is relative to the best non-\ULTRAG baseline in each column.}
\label{tab:cwq_results}
\begin{tabular}{lcccc}
\toprule
\textbf{Model} & \textbf{Hits} & \textbf{Recall} & \textbf{Precision} & \textbf{F1} \\
\midrule
GPT-5 & 60.5 & 53.2 & 56.3 & 52.7 \\
GCR & 69.3 & 62.1 & 66.3 & 62.2 \\
GNN-RAG & 75.9 & 70.3 & \textbf{71.7} & \textbf{69.2} \\
RoG & 75.8 & \textbf{70.9} & 70.5 & 68.8 \\
SubgraphRAG & \textbf{76.4} & 70.6 & 71.1 & 68.8 \\
\ULTRAGnospace-OTS & 73.7 & 68.8 & 68.0 & 66.3 \\
$\Delta$ & -2.7 & -2.1 & -3.7 & -2.9 \\
\bottomrule
\end{tabular}
\end{table}

We do not use WebQSP and CWQ as primary benchmarks because of their documented quality issues. These results nevertheless show that \ULTRAGnospace-OTS remains competitive on widely used legacy benchmarks, despite being fully inductive and requiring no dataset-specific training.

\section{Retrieval-inclusive Runtime}
\label{app:retrieval_runtime}

\begin{table}[H]
\centering
\caption{Runtime audit after including PPR sampling and FAISS retrieval. Times are average seconds per query; speed-up compares \ULTRAGnospace-OTS to the fastest baseline under the same accounting.}
\label{tab:retrieval_runtime}
\begin{tabular}{lccc}
\toprule
\textbf{Dataset} & \textbf{\ULTRAGnospace-OTS} & \textbf{Best baseline} & \textbf{Speed-up} \\
\midrule
WikiKG2 & 0.14s & 1.94s & 14.1$\times$ \\
Wikidata & 1.17s & 5.02s & 4.3$\times$ \\
\bottomrule
\end{tabular}
\end{table}

The main efficiency table isolates model-side non-API runtime on the extracted subgraphs. Adding the retrieval stages changes \ULTRAGnospace-OTS runtime by only $+0.04$s on WikiKG2 and $+1.02$s on Wikidata, so the full pipeline remains faster than the strongest baselines. This also clarifies the scaling claim: the pipeline retrieves from Wikidata-scale graphs, while the neural executor operates on PPR-selected subgraphs to keep query execution computationally tractable.





\end{document}